\documentclass[journal=jctcce,manuscript=article]{achemso}

\usepackage[version=3]{mhchem} 
\usepackage[utf8]{inputenc}
\usepackage[T1]{fontenc}
\usepackage{amsmath, amssymb, amsfonts, amsthm}
\usepackage{mathtools}
\usepackage{bm}
\usepackage{geometry}
\usepackage{graphicx}
\usepackage{subcaption}
\usepackage{listings}
\usepackage{xcolor}
\usepackage{booktabs}
\usepackage{longtable}
\usepackage{algorithm}
\usepackage{makecell}
\usepackage{algpseudocode}
\usepackage{tikz}
\usetikzlibrary{arrows.meta, positioning, shapes.geometric, calc}
\usepackage{fancyhdr}
\usepackage{titlesec}
\usepackage{enumitem}
\usepackage{caption}
\usepackage{subcaption}
\usepackage{float}
\usepackage{tabularx}
\usepackage{tcolorbox}
\usepackage{textcomp}
\tcbuselibrary{skins, breakable}
\usepackage{hyperref}
\hypersetup{
    colorlinks=true,
    linkcolor=blue!70!black,
    citecolor=green!50!black,
    urlcolor=blue!60!black
}
\usepackage{xurl}
\usepackage{array}

\usepackage{xcolor}   
\usepackage{soul}     

\DeclareUrlCommand{\code}{\urlstyle{tt}}
\newcolumntype{Y}{>{\raggedright\arraybackslash}X}

\definecolor{vibblue}{HTML}{1B3A5C}
\definecolor{viblightblue}{HTML}{D6E4F0}
\definecolor{vibmidblue}{HTML}{2E5F8A}
\definecolor{vibaccent}{HTML}{4A90D9}
\definecolor{vibwarn}{HTML}{C05050}

\newtcolorbox{keybox}[1][]{%
  enhanced, breakable,
  colback=viblightblue!30,
  colframe=vibblue,
  boxrule=1pt, arc=4pt,
  left=6pt, right=6pt, top=4pt, bottom=4pt,
  #1}

\newcommand{\bQ}{\mathbf{Q}}
\newcommand{\bn}{\mathbf{n}}
\newcommand{\bmv}{\mathbf{m}}
\newcommand{\Hvib}{\hat{H}_{\mathrm{vib}}}
\newcommand{\vibra}{\textsc{ViBra}}
\newcommand{\MI}[1]{\mathcal{I}_{#1}}

\newcommand{\Sexcl}[1]{S_{\mathrm{all}\setminus #1}}

\author{Raphael F. Ligorio}
\email{raphafe96@gmail.com}
\author{Marco A. Barroca}
\author{Alan Duriez}

\affiliation[Unknown University]
{Centro Brasileiro de Pesquisas Físicas (CBPF), 22290-180, Rio de Janeiro, RJ, Brasil}

\author{Mathias B. Steiner}

\email{mathias@kunumi.com}

\affiliation[Unknown University]
{Instituto Kunumi, 30130-138, Belo Horizonte, MG, Brasil}

\title{\vibra: Configuration Interaction for Anharmonic
Vibrational Spectroscopy and Quantum-Sampled Configuration Spaces}

\abbreviations{IR, VSCF, VCI, Spectroscopy}

\begin{document}

\begin{tocentry}
\centering
\vfill
\begin{tikzpicture}
    \node[inner sep=0] (img) {\includegraphics[width=7.6cm, height=4.6cm]{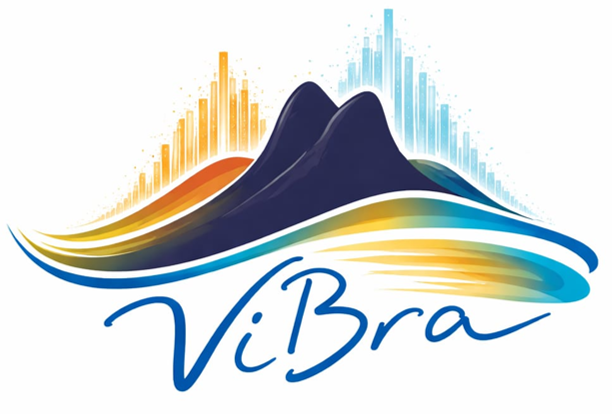}};
    \node[anchor=north, yshift=-0.1cm, text=black, font=\sffamily\fontsize{8}{10}\selectfont, align=center, text width=7.6cm] at (img.south) {\textbf{\vibra{}} -- a program for VSCF/VCI. The logo is inspired on the sunset at Morro Dois Irm\~aos, a classic landscape postcard from Rio de Janeiro, resembling two overlapping Gaussian peaks.};
\end{tikzpicture}
\vfill
\end{tocentry}
\begin{abstract}

Quantum-centric workflows are a promising route to improving the accuracy of property predictions in computational chemistry and materials science. By integrating quantum sampling algorithms with classical solvers, electronic structure calculations have recently demonstrated their potential even on noisy intermediate-scale quantum devices. In principle, the method of Vibrational Configuration Interaction (VCI) is suitable for integration with quantum sampling algorithms as well. However, demonstrations of computational workflows for quantum-centric, vibrational property predictions are still lacking. Here, we introduce a methodology for performing anharmonic vibrational structure calculations that can be deployed in a hybrid, quantum-classical mode. Starting from a quartic force field, the approach combines a Vibrational Self-Consistent Field (VSCF) with VCI in either Full, Selected (S-VCI), or Symmetry-Adapted (SA-VCI) mode. In S-VCI, an Epstein--Nesbet perturbative screening significantly reduces the configuration space while retaining high predictive accuracy. A state-list input enables the integration of externally generated vibrational configurations as a seed space. As a proof-of-concept, we demonstrate a hybrid, quantum-classical computational workflow, in which a quantum sampling algorithm provides the seed. Our vibrational wave function analysis package \vibra{}, equipped with a graphical interface, is available at \url{https://github.com/raphafe96/ViBra}.

\end{abstract}

\section{Introduction}
\label{sec:intro}

Vibrational spectroscopy provides a direct window into molecular structure and dynamics.
The positions and intensities of fundamental, overtone, and combination bands encode
information about bond strengths, mode couplings, and the underlying potential energy surface (PES).
Accurate theoretical prediction of these observables therefore constitutes a stringent test
of electronic structure methods and is indispensable for the assignment of experimental spectra.

Within the Born--Oppenheimer approximation, the nuclear vibrational problem can be formulated
as the solution of the vibrational Schrödinger equation in mass-weighted normal coordinates
$\bQ = (Q_1, \dots, Q_M)$, where $M = 3N_{\text{at}}-6$ or $M = 3N_{\text{at}}-5$ for linear molecules is the number of vibrational degrees of freedom. The simplest harmonic approach approximation uses a quadratic PES and diagonalizes the mass-weighted Hessian to yield independent normal modes. However, this approximation fails to reproduce
experimental fundamental frequencies, cannot describe overtones and
combination bands, and gives vanishing intensities for these transitions within the
double-harmonic (mechanical and electrical) approximation~\cite{barone2005}.
The accurate treatment of mechanical anharmonicity requires the inclusion of at least cubic and quartic terms in the PES expansion~\cite{barone2005,christiansen2007}.

Two broad classes of methods have been developed for the anharmonic vibrational problem.
\textbf{Perturbative} approaches, exemplified by second-order vibrational perturbation theory
(VPT2)~\cite{barone2005,bloino2015}, are computationally efficient and have been widely deployed
in quantum chemistry codes such as Gaussian~\cite{barone2005} and ORCA~\cite{neese2020,neese2022}.
However, VPT2 breaks down in the presence of Fermi and Darling--Dennison resonances,
requiring ad hoc treatments that can compromise accuracy.

\textbf{Variational} methods, by contrast, are systematically improvable and are robust
in the presence of resonances. The Vibrational Self-Consistent Field (VSCF) approach,
introduced independently by Bowman~\cite{bowman1978} and by Gerber and Ratner~\cite{gerber1979},
provides a mean-field description in which each normal mode evolves in the average potential
of all other modes. To recover the correlation missed by the mean-field treatment, VSCF
is typically followed by Vibrational Configuration Interaction (VCI), where the wavefunction
is expanded in a basis of Hartree products of VSCF-optimised one-mode functions (modals)
and the Hamiltonian is diagonalised in this space~\cite{christiansen2004,bowman2008,christiansen2007}.

Over the past three decades, several powerful VCI implementations have been developed.
The \textbf{MULTIMODE} code, developed by Bowman and co-workers~\cite{bowman2008},
uses an $n$-mode representation of the potential in normal coordinates with a hierarchical
coupling scheme and has been applied to molecules with up to dozens of atoms.
The \textbf{MOLPRO} package features state-specific configuration-selective VCI (cs-VCI)
algorithms pioneered by Rauhut and co-workers~\cite{rauhut2007,neff2009}, which use
perturbative estimates to efficiently screen the correlation space.
More recently, Mathea and collaborators have introduced advanced screening techniques
and efficient handling of rovibrational angular momentum terms that substantially accelerate
these calculations~\cite{mathea2021,mathea2022}.

The \textbf{CRYSTAL} program, developed by Erba, Maul, Dovesi, and co-workers,
has extended VSCF and VCI methods to periodic systems (solids), implementing a fully
automated scheme for constructing the anharmonic PES from density functional theory
(DFT) calculations and computing anharmonic vibrational states of solids~\cite{erba2019, erba2019a}.
A subsequent extension by Carbonnière et al. additionally enabled the calculation of
anharmonic infrared and Raman intensities for periodic systems~\cite{carbonniere2020}.

In the open-source domain, the \textbf{PyVCI} package by Sibaev and Crittenden~\cite{sibaev2016}
provides a flexible Python-based VCI implementation with computationally intensive routines
written in C via Cython. The \textbf{DVCI} (Dual Vibration Configuration Interaction)
method by Garnier et al.~\cite{garnier2019} employs an efficient factorisation of the
Hamiltonian to target specific spectral transitions with minimal memory usage, while the
\textbf{A-VCI} (Adaptive VCI) algorithm ~\cite{Garnier2016, Odunlami2017} iteratively
constructs nested bases guided by a posteriori error estimates.

A significant recent advance has been the development of \textbf{Vibrational Heat-Bath
Configuration Interaction (VHCI)}. Introduced independently by Fetherolf and Berkelbach
in 2021~\cite{fetherolf2021} and by Bhatty and Brorsen in the same year~\cite{bhatty2021},
VHCI adapts the heat-bath CI algorithm from electronic structure theory to the vibrational
domain. It uses a simple, pre-sorted list of anharmonic force constants to efficiently
identify important basis states, achieving near-full-VCI accuracy at a fraction of the
computational cost. A subsequent extension incorporated VSCF modals as the one-mode
basis and semistochastic perturbation theory, further improving convergence~\cite{tran2023}.

Despite this wealth of methodology, several challenges remain: the exponential scaling
of the VCI basis with $M$, the accurate treatment of transition intensities beyond the
linear dipole surface, and the need for user-friendly software that bridges computation
and visualisation. We therefore designed \textbf{\vibra} to address these challenges
through a combination of:

\begin{itemize}
  \item \textbf{Selected CI (S-VCI):} an Epstein--Nesbet perturbation-theory-driven
    configuration selection that significantly reduces the VCI space while preserving
    accuracy for low-lying states (Section~\ref{sec:SCI});
  \item \textbf{Symmetry-Adapted VCI:} exact block-diagonalisation of the Hamiltonian
    for molecules in Abelian point groups with one-dimensional irreducible representations
    ($C_1$, $C_s$, $C_i$, $C_2$, $C_{2h}$, $C_{2v}$, $D_2$, $D_{2h}$), providing irrep
    labels and reducing diagonalisation cost by up to a factor of $n_\Gamma^2$
    (Section~\ref{sec:symmetry}), where $n_\Gamma$ is the number of symmetry elements of a given Abelian point group;
  \item \textbf{Treatment of electrical anharmonicity:} all VCI-level transition
    dipoles include both first- and second-order dipole moment derivatives, essential
    for overtone and combination band intensities and for the correct redistribution
    of oscillator strength through Fermi resonances (Section~\ref{sec:dipole});
  \item \textbf{Graphical interface and interactive viewer:} a Python/CustomTkinter GUI
    for input preparation and job execution, and a Matplotlib-based viewer with
    temperature-dependent spectra, experimental overlay, baseline correction, and
    3D normal-mode animation via 3Dmol.js (Section~\ref{sec:GUI}).
\end{itemize}

A further motivation for the development of \vibra{} is the growing interest in quantum algorithms for vibrational structure and spectroscopy. In quantum chemistry and materials science, recent quantum-centric workflows such as sample-based quantum diagonalization (SQD) and sample-based Krylov quantum diagonalization (SKQD) have highlighted a practical near-term paradigm in which quantum processors are used as sampling devices, while compact Hamiltonian construction, diagonalization, and observable evaluation are performed classically~\cite{RobledoMoreno2025SQD,Yu2025SKQD}. More broadly, quantum-centric supercomputing has been proposed as a route for integrating quantum processors with high-performance classical resources in materials-science workflows~\cite{Alexeev2024QCSCMaterials}. In contrast to electronic-structure applications, however, quantum algorithms for anharmonic vibrational dynamics remain comparatively underdeveloped, despite recent progress in quantum algorithms for vibronic spectra, vibrational structure calculations, and Trotterized simulation of vibrational Hamiltonians~\cite{Sawaya2019VibronicSpectra,Ollitrault2020VibrationalStructure,Malpathak2025TrotterVibrational}. 

The selected-VCI infrastructure implemented in \vibra{} is well aligned with
this sampling-based perspective because it can serve as the classical VSCI
engine after a quantum-generated seed space has been obtained. A quantum
sampling algorithm proposes a compact set of vibrational configurations,
while \vibra{} can either construct and diagonalize the Hamiltonian directly
in that sampled space or enlarge it through EN-PT2-guided selection before
variational rediagonalization. In both cases, \vibra{} evaluates transition
observables and generates anharmonic vibrational spectra from the final
configuration space. Thus, \vibra{} provides both the projected-VCI
post-processing layer and an optional classical refinement stage for
quantum-seeded vibrational configuration spaces.

\section{Theoretical Framework}

The potential energy surface (PES) governing nuclear motion can be represented
in various forms, including numerical grids or analytical expansions. In
\vibra{}, we have opted for a Taylor expansion of the PES in terms of the
normal coordinates, truncated at fourth order~\cite{erba2019a}. This quartic
force field representation provides a compact, analytically tractable form
that enables efficient evaluation of all required matrix elements in closed
form, since low-order polynomial terms in $Q_i$ have simple analytic matrix
elements in the harmonic-oscillator basis.
Furthermore, it allows for a clean separation of the potential into harmonic,
diagonal anharmonic, and coupling contributions, which is essential for the
VSCF and VCI formulations described below. Grid-based representations such as
the discrete variable representation (DVR)~\cite{light2000}, while more
flexible and not restricted to a low-order polynomial form, require the PES
to be evaluated on a dense multidimensional grid whose size grows
exponentially with the number of modes, and they do not offer a natural
analytic separation into diagonal and coupling terms; this makes them less
suitable for the sparsity-driven algorithms (Section~\ref{sec:optim}) that
\vibra{} relies on. 

\label{sec:theory}

\subsection{Vibrational Hamiltonian and Quartic Force Field}
\label{sec:Hamiltonian}

For a non-linear molecule with $N_{\mathrm{at}}$ atoms, mass-weighted normal coordinates
$\bQ = (Q_1,\dots,Q_M)$ are obtained by diagonalizing the mass-weighted Cartesian Hessian,
where $M = 3N_{\mathrm{at}}-6$ is the number of vibrational degrees of freedom.
Neglecting rotation--vibration coupling ($J=0$) and the Watson correction terms that are
small for semi-rigid molecules, the vibrational Hamiltonian in atomic units reads

\begin{equation}
\Hvib = -\frac{1}{2}\sum_{i=1}^{M}\frac{\partial^{2}}{\partial Q_i^{2}} + V(\bQ).
\label{eq:H}
\end{equation}

The potential energy surface $V(\bQ)$ is expanded in a Taylor series about the equilibrium
geometry, here up to fourth order:

\begin{equation}
V(\bQ) = \frac{1}{2}\sum_{i=1}^{M} \omega_i^{2} Q_i^{2}
        + \frac{1}{6}\sum_{i=1}^{M}\sum_{j=1}^{M}\sum_{k=1}^{M}
            \phi_{ijk}\, Q_i Q_j Q_k
        + \frac{1}{24}\sum_{i=1}^{M}\sum_{j=1}^{M}\sum_{k=1}^{M}\sum_{l=1}^{M}
            \phi_{ijkl}\, Q_i Q_j Q_k Q_l .
\label{eq:qff_full}
\end{equation}

The harmonic frequencies $\omega_i$ and the cubic and quartic force constants
$\phi_{ijk}$ and $\phi_{ijkl}$ are provided exclusively in the ORCA VPT2 output format
(\texttt{.vpt2})~\cite{neese2020,neese2022}. This format was chosen because it provides all necessary information, including harmonic frequencies, cubic and quartic force constants with their permutational degeneracy factors, normal-coordinate transformation matrices, and first and second dipole derivatives, in a self-contained, well-documented text structure. Notably, ORCA is an free quantum chemistry package, making the entire computational workflow from force constant evaluation to anharmonic vibrational analysis fully freely accessible. Users of other quantum chemistry packages can convert their outputs to the equivalent format using as reference the ORCA output files provided in Support Information. 

The tensors are totally symmetric under
index permutation; only the upper-triangular components are stored, and each distinct
term carries a permutational degeneracy factor. For cubic terms with
$i \le j \le k$, the weight $g$ is 1 (all indices equal), 3 (two equal), or 6 (all distinct).
For quartic terms $i \le j \le k \le l$, the weight is 1 (all equal), 4 (three equal),
6 (two doubles), 12 (one double, two singles), or 24 (all distinct). The code stores the
effective potential in terms of these weighted coefficients as

\begin{align}
V(\bQ) &= \sum_i \Bigl(\tfrac12 \omega_i^2 Q_i^2
                     + \tilde\phi_{iii}\,Q_i^3
                     + \tilde\phi_{iiii}\,Q_i^4\Bigr)
        + \sum_{\substack{i<j\leq k}} \tilde\phi_{ijk}\,Q_i Q_j Q_k
        + \sum_{\substack{i<j\leq k\leq l}} \tilde\phi_{ijkl}\,Q_i Q_j Q_k Q_l .
\label{eq:qff_weighted}
\end{align}


The coefficients $\tilde\phi_{\cdots}$ incorporate the original force constants
and the appropriate degeneracy factors. Internally, a separate diagonal potential
$V_i^{\mathrm{diag}}(Q_i) = \tfrac12 \omega_i^2 Q_i^2 + \tilde\phi_{iii} Q_i^3
+ \tilde\phi_{iiii} Q_i^4$ is identified for each mode, and the coupling potential
$V_c$ groups all terms involving at least two distinct mode indices.

\subsection{One-Mode Basis: Harmonic Oscillator}
\label{sec:HO}

The primitive one-mode basis consists of harmonic oscillator eigenfunctions
$\phi_n(Q_i)$ with frequency $\omega_i$. Their matrix elements
\begin{equation}
m_{p}^{(\mu,\nu)} \equiv \langle\phi_{\mu} | Q_i^{p} | \phi_{\nu} \rangle
\end{equation}
for $p = 0,\dots,4$ are evaluated analytically using ladder-operator algebra.
The kinetic energy matrix elements are computed from the second-derivative operator:
\begin{equation}
\langle\phi_{\mu} | \hat{T}_i | \phi_{\nu} \rangle
= -\frac12 \langle\phi_{\mu} | \frac{\partial^{2}}{\partial Q_i^{2}} | \phi_{\nu} \rangle.
\end{equation}
A fundamental selection rule governs these integrals:
$\langle\phi_{\mu} | Q_i^{p} | \phi_{\nu} \rangle = 0$ whenever
$|\mu - \nu| > p$. All required matrix elements up to $p = 4$ are
pre-computed by the \texttt{compute\_integrals} module and stored in memory
for efficient look-up.

\subsection{VSCF Method and Modal Basis}
\label{sec:VSCF}

The VSCF wavefunction is the Hartree product
\begin{equation}
\Psi_{\bn}(\bQ) = \prod_{i=1}^{M} \varphi_{n_i}^{(i)}(Q_i),
\label{eq:prod}
\end{equation}
where $\bn = (n_1,\dots,n_M)$ specifies the quantum state of each mode.
Each one-mode function (modal) is expanded in the HO basis:

\begin{equation}
\varphi_{n_i}^{(i)}(Q_i) = \sum_{\mu=1}^{N_{\mathrm{exp}}}
  C_{\mu,n_i}^{(i)}\,\phi_{\mu-1}(Q_i).
\label{eq:modal_exp}
\vspace{-2mm}
\end{equation}

Here $N_{\mathrm{exp}}$ is the number of HO basis functions per mode (user-supplied
keyword \texttt{NEXPAN}; typical values range from 8 to 16).

Requiring the energy functional $E_{\bn} = \langle\Psi_{\bn}|\Hvib|\Psi_{\bn}\rangle$
to be stationary with respect to variations of each modal under the constraint
$\langle\varphi_{n_i}^{(i)}|\varphi_{n_i}^{(i)}\rangle = 1$ yields $M$ coupled
effective one-mode equations:

\begin{equation}
\hat h_i^{\mathrm{eff}}\;\varphi_{n_i}^{(i)}(Q_i) =
  \varepsilon_{n_i}^{(i)}\,\varphi_{n_i}^{(i)}(Q_i),
\label{eq:modal_eq}
\end{equation}

where $\varepsilon_{n_i}^{(i)}$ acts as the Lagrange multiplier enforcing normalisation
and has the physical interpretation of a modal energy. The effective Hamiltonian

\begin{equation}
\hat h_i^{\mathrm{eff}} = \hat T_i + V_i^{\mathrm{diag}}(Q_i)
   + \langle V_c\rangle_{j\neq i}
\label{eq:heff}
\end{equation}

contains the kinetic energy of mode $i$, its diagonal potential, and the mean-field
average of the coupling potential over all other modes:
\begin{equation}
\langle V_c\rangle_{j\neq i}
= \Big\langle\prod_{j\neq i} \varphi_{n_j}^{(j)} \Big|
  V_c(\bQ)
  \Big|\prod_{j\neq i} \varphi_{n_j}^{(j)} \Big\rangle.
\end{equation}

Because $V_c$ is a polynomial of degree four in the normal coordinates, the
mean-field average is itself a polynomial in $Q_i$, so that

\begin{equation}
\hat h_i^{\mathrm{eff}} = \hat T_i + \sum_{p=0}^{4} X_i^{(p)}\,Q_i^{p}.
\label{eq:heff_poly}
\end{equation}

The coefficients $X_i^{(p)}$ are real numbers that depend on the force constants
and on the current expectation values $\langle\varphi_{n_j}^{(j)}|Q_j^{c}|\varphi_{n_j}^{(j)}\rangle$
of all modes $j \neq i$. They are updated at each self-consistent iteration.

Substituting the modal expansion~\eqref{eq:modal_exp} into~\eqref{eq:modal_eq} and
projecting onto $\langle\phi_{\nu-1}|$ yields an $N_{\mathrm{exp}} \times N_{\mathrm{exp}}$
real symmetric matrix eigenvalue problem:

\begin{equation}
\sum_{\mu=1}^{N_{\mathrm{exp}}} H_{\nu\mu}^{(i)}\, C_{\mu,n_i}^{(i)}
= \varepsilon_{n_i}^{(i)}\,C_{\nu,n_i}^{(i)},
\label{eq:H_i}
\end{equation}

with matrix elements

\begin{equation}
H_{\nu\mu}^{(i)} = \langle\phi_{\nu-1}|\hat T_i|\phi_{\mu-1}\rangle
                 + \sum_{p=0}^{4} X_i^{(p)}\,
                   \langle\phi_{\nu-1}|Q_i^{p}|\phi_{\mu-1}\rangle.
\label{eq:H_i_elements}
\end{equation}

The $n_i$-th eigenvector gives the updated coefficients $C_{\mu,n_i}^{(i)}$.
The VSCF equations are solved iteratively (Algorithm~\ref{alg:VSCF}) until the energy
change between successive iterations falls below $10^{-\texttt{CVGSCF}}$ cm$^{-1}$.
The ground-state modals are stored for use as the VCI one-mode basis.

Because each coupling term in $V_c$ is counted in the effective Hamiltonian of every
mode (either as an operator or as a constant shift $X_i^{(0)}$), the sum of modal
energies $\sum_i \varepsilon_{n_i}^{(i)}$ overcounts the coupling energy. The correct
VSCF total energy is therefore

\begin{equation}
E_{\mathrm{VSCF}} = \sum_{i=1}^{M} \varepsilon_{n_i}^{(i)}
                  - (M-1)\,\bar V_c,
\label{eq:Evscf}
\end{equation}

where the correlation potential

\begin{equation}
\bar V_c = \Big\langle\prod_{i=1}^{M}\varphi_{n_i}^{(i)} \Big|
           V_c(\bQ)
           \Big|\prod_{i=1}^{M}\varphi_{n_i}^{(i)} \Big\rangle
\end{equation}

is the fully averaged coupling potential. This correction ensures that each coupling
term appears exactly once in the total energy.

\begin{algorithm}[H]
\caption{VSCF Self-Consistent Field Iteration}
\label{alg:VSCF}
\begin{algorithmic}[1]
\State Initialize modal coefficients $C_{\mu,n_i}^{(i)}$ to small uniform values for all modes
\Repeat
  \State Precompute all one-mode expectation values
          $\langle\varphi_{n_j}^{(j)}|Q_j^{c}|\varphi_{n_j}^{(j)}\rangle$
          for $c=0,\dots,4$ (overlap cache)
  \For{$i = 1$ to $M$}
    \State Construct $X_i^{(p)}$ ($p=0,\dots,4$) from force constants and current overlaps
    \State Build $H_{\nu\mu}^{(i)}$ via Eq.~\eqref{eq:H_i_elements}
    \State Diagonalize $\mathbf{H}^{(i)}$ (LAPACK \texttt{DSYEVD});
           retain $C_{\mu,n_i}^{(i)}$ from the $n_i$-th eigenvector
  \EndFor
  \State Compute $\bar V_c$ as the fully contracted coupling potential
  \State $E_{\mathrm{VSCF}} \gets \sum_i\varepsilon_{n_i}^{(i)} - (M-1)\bar V_c$
\Until{$|E_{\mathrm{VSCF}} - E_{\mathrm{old}}| < 10^{-\texttt{CVGSCF}}$ cm$^{-1}$}
\end{algorithmic}
\end{algorithm}

\subsection{VCI Expansion and Precomputed Modal Integrals}
\label{sec:VCI}

The VCI expansion can be built using either the primitive harmonic oscillator basis or the ground-state VSCF modals. The VSCF modals yield faster variational convergence of the VCI energies with respect to increasing the maximum total quanta, as they already account for the average effect of anharmonic couplings. This is achieved without additional computational overhead, since all required modal integrals are precomputed and stored (see Section \ref{sec:optim}). The VSCF-based VCI approach is therefore recommended for routine anharmonic vibrational calculations and thus employed in \vibra{}. As a result, the wavefunction is a linear combination of Hartree products
built from the \emph{ground-state} VSCF modals:

\begin{equation}
|\Psi_I\rangle = \sum_{\bn\in\mathcal{S}} c_{\bn}^{(I)}\;|\Phi_{\bn}\rangle,
\qquad
|\Phi_{\bn}\rangle = \prod_{i=1}^{M} |\varphi_{n_i}^{(i)}\rangle.
\label{eq:vci_ansatz}
\end{equation}

The configuration space $\mathcal{S}$ is truncated by a maximum total quantum number
$N_q$ (keyword \texttt{NQUANT}) such that $\sum_{i=1}^{M} n_i \le N_q$.
Using a single fixed set of ground-state modals ensures orthonormality of the
configuration basis. A critical component enabling efficient VCI is the precomputed
modal integral table:

\begin{equation}
\MI{i}(v,v';p) \equiv \langle\varphi_v^{(i)}|\;Q_i^p\;|\varphi_{v'}^{(i)}\rangle
= \sum_{\mu,\nu=1}^{N_{\mathrm{exp}}}
  C_{\mu,v}^{(i)} C_{\nu,v'}^{(i)}\,
  \langle\phi_{\mu-1}|\;Q_i^{p}\;|\phi_{\nu-1}\rangle,
\label{eq:modal_int}
\end{equation}

with $p=0,\dots,5$ (where $p=5$ stores kinetic energy matrix elements in the
modal basis). The table is symmetric in its first two arguments and occupies
$M \times (N_q+1)^2 \times 6$ double-precision entries---typically a few tens
of megabytes. Once computed, every one-mode integral required for the VCI
Hamiltonian construction becomes an $O(1)$ look-up, eliminating the need for
the $O(N_{\mathrm{exp}}^2)$ double sum otherwise required for each evaluation.

The Hamiltonian matrix element $H_{\bmv\bn} \equiv \langle\Phi_{\bmv}|\Hvib|\Phi_{\bn}\rangle$
is assembled from three types of contribution:

\paragraph{One-body terms} collect all diagonal anharmonic corrections for
each mode $i$:
\begin{equation}
H^{(1)} = \sum_{i=1}^{M}
\Bigl[
  \MI{i}(m_i,n_i;5)
  + \tilde\phi_{iiii}\,\MI{i}(m_i,n_i;4)
  + \tilde\phi_{iii}\,\MI{i}(m_i,n_i;3)
  + \tfrac{\omega_i}{2}\,\MI{i}(m_i,n_i;2)
\Bigr]\,\Sexcl{i}.
\label{eq:H1b}
\end{equation}

\paragraph{Cubic coupling terms} with unique modes $\{s_1,s_2,s_3\}$ and
multiplicities $\{c_1,c_2,c_3\}$ contribute
\begin{equation}
H^{(3,\text{term})}
= \tilde\phi_{\text{term}}\;
  \prod_{r=1}^{3} \MI{s_r}(m_{s_r},n_{s_r};c_r)\;
  \Sexcl{\{s_1,s_2,s_3\}} .
\label{eq:H3}
\end{equation}

\paragraph{Quartic coupling terms} are treated analogously with up to four unique modes.

In the above, $\Sexcl{S}$ denotes the product of spectator-mode overlaps:
\begin{equation}
\Sexcl{S} = \prod_{k \notin S} \MI{k}(m_k,n_k;0).
\label{eq:Sexcl}
\end{equation}

A fundamental selection rule emerges: because $\MI{k}(m_k,n_k;0) = \delta_{m_k,n_k}$
(modal orthonormality), a potential term can couple configurations $\bmv$ and $\bn$
only if it involves \emph{all} modes in which the two configurations differ.
Consequently, two VCI configurations can be coupled only if they differ in at most
four modes (the maximum number of distinct indices in a quartic term). This sparsity
is exploited through a precomputed pair list and an inverted potential index.

The Hamiltonian is diagonalised using LAPACK routines: \texttt{DSYEVD} (divide-and-conquer)
for full eigenvalue spectra, and \texttt{DSYEVR} (relatively robust representations)
for partial spectra when only the lowest $N_{\mathrm{state}}$ eigenvalues are required.

\subsection{Selected Vibrational CI (S-VCI)}
\label{sec:SCI}

Full VCI becomes prohibitive for large $M$ and $N_q$ since
$|\mathcal{S}| \sim \binom{M+N_q}{N_q}$. \vibra's SCI algorithm addresses this
by constructing a compact active space through Epstein--Nesbet perturbation theory~\cite{epstein1926,nesbet1955}
screening (EN-PT).

\subsubsection{Reference Space}
\label{sec:cisd_ref}

The reference space $\mathcal{S}_{\mathrm{ref}}$ is set by the user through the
\texttt{MAXSCI} keyword,
\begin{center}
\texttt{MAXSCI} $N_{\mathrm{sel}}$ \texttt{[auto\,|\,list]} \texttt{[}$n_{\mathrm{ref}}$\texttt{]}
\end{center}
where $N_{\mathrm{sel}}$ is the number of external configurations retained per
reference state (Section~\ref{sec:topN}), the second argument selects how the
reference space is built, and the third, optional, argument sets the truncation
order $n_{\mathrm{ref}}$, used only when the \texttt{auto} option is chosen.

In \texttt{auto} mode the reference space comprises all configurations whose total
excitation does not exceed $n_{\mathrm{ref}}$,
\begin{equation}
\mathcal{S}_{\mathrm{ref}} = \Bigl\{\, \bn \in \mathcal{S} \;:\;
\textstyle\sum_i n_i \le n_{\mathrm{ref}} \Bigr\}.
\label{eq:ref_space_auto}
\end{equation}
For $n_{\mathrm{ref}}=2$ (ground state, one-mode singles $n_i=1$, and doubles
$n_i=2$ or $n_i=n_j=1$ pairs) this has dimension $1 + 2M + \binom{M}{2}$, linear in
$M$ for the singles and quadratic for the doubles; larger values of $n_{\mathrm{ref}}$
extend the reference to include triples, quadruples, and so on, at the cost of a
larger reference Hamiltonian.

In \texttt{list} mode the reference space is instead supplied explicitly by the
user through a plain-text file, \texttt{list\_states.txt}, whose first line gives
the number of reference configurations, followed by one configuration (a length-$M$
vector of mode quanta $n_i$) per line. This allows the user to target
configurations of specific chemical or spectroscopic interest, independent of any
excitation-order truncation. In either case, the Hamiltonian in the resulting reference space $\mathcal{S}_{\mathrm{ref}}$
is built and diagonalised exactly, yielding reference energies $E_I^{(0)}$ and
eigenvectors
$|\Psi_I^{(0)}\rangle = \sum_{\bn\in\mathcal{S}_{\mathrm{ref}}}
c_{\bn}^{(I)}|\Phi_{\bn}\rangle$, for $I=1,\dots,N_{\mathrm{state}}$, where
$N_{\mathrm{state}}$ here is the number of vibrational reference states.

\subsubsection{Epstein--Nesbet PT2 Screening}
\label{sec:enpt2}

For each external configuration $|\Phi_\alpha\rangle \notin \mathcal{S}_{\mathrm{ref}}$
and each reference state $I$ to be converged, the Epstein--Nesbet second-order
energy correction is computed:
\begin{equation}
e_\alpha^{(I)} =
\frac{\bigl|
   \sum_{\bn\in\mathcal{S}_{\mathrm{ref}}} c_{\bn}^{(I)} H_{\alpha\bn}
   \bigr|^2}
     {E_I^{(0)} - H_{\alpha\alpha}} .
\label{eq:enpt2}
\end{equation}

The denominator $H_{\alpha\alpha}$ is the diagonal Hamiltonian element of the
external configuration itself, $H_{\alpha\alpha} = \langle\Phi_\alpha|H|\Phi_\alpha\rangle$.
This is the defining feature of the Epstein--Nesbet partitioning of the
Hamiltonian: rather than approximating a
configuration's zeroth-order energy as a sum of independent-mode (harmonic)
energies, as would be done in a M{\o}ller--Plesset-type partitioning, the exact
diagonal matrix element---already including diagonal anharmonic
corrections---is used directly. This makes the energy denominator in
Eq.~\eqref{eq:enpt2} a tighter and more physically faithful estimate for each
configuration, at the cost of no additional computation beyond evaluating
$H_{\alpha\alpha}$. The numerator in Eq.~\eqref{eq:enpt2} instead couples
$|\Phi_\alpha\rangle$ to every reference configuration $\bn$ that differs by at
most four quanta; this coupling is evaluated efficiently using the inverted
potential index (Section~\ref{sec:optim}).

\subsubsection{Top-\texorpdfstring{$N$}{N} Selection and Active Space Construction}
\label{sec:topN}

For each reference state $I$, the $N_{\mathrm{sel}}$ external configurations with
the largest $|e_\alpha^{(I)}|$ are retained, where $N_{\mathrm{sel}}$ is the first
argument of the \texttt{MAXSCI} keyword. A sorted insertion procedure maintains
the top-$N$ candidates for each state as the external space is scanned once,
avoiding a full sort of all PT2 contributions. The final active space is the union
\begin{equation}
\mathcal{S}_{\mathrm{active}} = \mathcal{S}_{\mathrm{ref}}
\;\cup\;\bigcup_{I=1}^{N_{\mathrm{state}}}
   \mathcal{S}_{\mathrm{sel}}^{(I)}.
\label{eq:active_space}
\end{equation}
Because a single external configuration may be selected by multiple reference
states, duplicate insertions are automatically avoided when the union is formed.
The full VCI Hamiltonian is then rebuilt in the final active space and
diagonalised using the same optimised kernel used for the reference space and for
the conventional full-VCI calculation.

For example, for the ethylene molecule, with 12 vibrational modes, a choice of 6
quanta, using $N_{\mathrm{sel}} = 100$ yields an active space corresponding to only
$32\%$ of the full VCI dimension, while the lowest eigenvalues agree with the full
VCI results within $1$--$5$ cm$^{-1}$. S-VCI is activated automatically when \texttt{MAXSCI} $N_{\mathrm{sel}} > 0$ and no
molecular symmetry is specified (i.e., \texttt{PGROUP} is \texttt{C1} or absent).
If symmetry is active, the code defaults to the symmetry-adapted full VCI.

\subsection{Symmetry-Adapted VCI (SA-VCI)}
\label{sec:symmetry}

\subsubsection{Motivation and Scope}

The vibrational Hamiltonian $\Hvib$ is invariant under every symmetry operation
of the molecular point group $\mathcal{G}$: $[\Hvib,\hat{R}] = 0$ for all
$\hat{R} \in \mathcal{G}$. Consequently, its matrix representation in a
symmetry-adapted basis is block-diagonal: matrix elements between configurations
of different irreducible representations (irreps) are identically zero.

\vibra{} exploits this for the eight Abelian point groups with exclusively
one-dimensional irreps: $C_1$, $C_s$, $C_i$, $C_2$, $C_{2h}$, $C_{2v}$, $D_2$,
$D_{2h}$. In these groups, all characters are $\pm 1$ and the irrep of a
Hartree product factorises as a direct product of one-mode irreps.

\subsubsection{Determination of Normal-Mode Irreps}
\label{sec:mode_irrep}

The irrep of each normal mode $k$ is determined by projecting its Cartesian
displacement vector $\mathbf{L}^{(k)}$ (a $3N_{\mathrm{at}}$-dimensional vector)
onto the symmetry operations in the principal-axis frame (PAF). The character
of mode $k$ under operation $\hat{R}$ is

\begin{equation}
\chi^{(k)}(\hat R) =
\frac{1}{\|\mathbf{L}^{(k)}\|^2}
\sum_{a=1}^{N_{\mathrm{at}}}
\sum_{\alpha,\beta} L_{\alpha,\hat R(a)}^{(k)}\,
   R_{\alpha\beta}\, L_{\beta,a}^{(k)}.
\label{eq:mode_char}
\end{equation}

Here $\mathbf{R}$ is the $3\times 3$ rotation/reflection matrix of the operation,
$\hat R(a)$ is the atom to which atom $a$ is mapped, and the double sum runs over
all three Cartesian components and all atoms. The sum includes both ``diagonal''
contributions (atoms mapped to themselves) and ``off-diagonal'' (permuted) ones,
both of which are essential for obtaining the correct $\pm1$ character when the
mode displacement is concentrated on permuted atoms.

The PAF is obtained by translating the molecule to its centre of mass, computing
the inertia tensor $I_{pq} = \sum_a m_a(\delta_{pq}|\mathbf{r}_a|^2 - r_{a,p}r_{a,q})$,
and diagonalising it (LAPACK \texttt{dsyev}). The rotated coordinates and displacement
vectors are used in Eq.~\eqref{eq:mode_char}.

The irrep $\Gamma_k$ of mode $k$ is identified by minimal $L^1$ distance between
the computed character vector $\boldsymbol{\chi}^{(k)}$ and the rows of the
point-group character table.

\subsubsection{Configuration Irreps and Block Diagonalisation}

The VSCF modals inherit the parity structure of the HO basis:
$\hat{R}\,\varphi_{n_i}^{(i)}(Q_i) = [\chi(\Gamma_i,\hat{R})]^{n_i}\,
\varphi_{n_i}^{(i)}(Q_i)$. Therefore the contribution of mode $i$ to the irrep
of a configuration is $\Gamma_i$ when $n_i$ is odd and $\Gamma_1$
(the totally symmetric irrep) when $n_i$ is even. The irrep of a configuration
is the direct product of all contributing odd-quanta mode irreps:

\begin{equation}
\Gamma(|\Phi_{\bn}\rangle) =
\bigotimes_{\substack{i=1\\ n_i\ \mathrm{odd}}}^{M} \Gamma(Q_i).
\label{eq:config_irrep}
\end{equation}

The configuration space partitions into $n_\Gamma$ blocks:
$\mathcal{S} = \bigcup_{\Gamma} \mathcal{B}_\Gamma$. Each block is diagonalised
independently, reducing the cost from $O(|\mathcal{S}|^3)$ to
$\sum_\Gamma O(|\mathcal{B}_\Gamma|^3)$. No approximation is introduced:
the symmetry-adapted VCI is exactly equivalent to the full VCI for the correct
group assignment.

\subsubsection{Diagnostic Output and User Verification}

The user specifies the point group via the \texttt{PGROUP} keyword. Because
\vibra{} cannot auto-detect the true symmetry, extensive diagnostic output is
printed to \texttt{stdout}: PAF coordinates, atom permutation tables, per-atom
contributions to each mode character, and the assigned irrep. A complete discussion of the consequences of incorrect group assignment is provided
in the Supporting Information.


\subsection{Transition Dipoles and Infrared Intensities}
\label{sec:dipole}

Infrared intensities are proportional to the squared transition dipole moment:

\begin{equation}
I_{0I} \propto \sum_{\alpha=x,y,z}
\bigl| \langle\Psi_0|\hat\mu_\alpha|\Psi_I\rangle \bigr|^2 .
\label{eq:intensity}
\end{equation}

The dipole operator is expanded to second order in normal coordinates:

\begin{equation}
\hat\mu_\alpha = \mu_\alpha^{(0)}
   + \sum_{i=1}^{M} \mu_\alpha^{(i)}\,Q_i
   + \frac12 \sum_{i=1}^{M}\sum_{j=1}^{M}
      \mu_\alpha^{(ij)}\,Q_i Q_j,
\label{eq:dipole_exp}
\end{equation}

where $\mu_\alpha^{(i)} \equiv \partial\mu_\alpha/\partial Q_i|_0$
and $\mu_\alpha^{(ij)} \equiv \partial^2\mu_\alpha/\partial Q_i\partial Q_j|_0$.

\subsubsection{Transformation of Dipole Derivatives from ORCA}
\label{sec:dipole_transform}

The ORCA VPT2 output provides the first derivatives in Cartesian coordinates,
$\partial\mu_\alpha/\partial x_k$. They are transformed to normal coordinates via

\begin{equation}
\frac{\partial\mu_\alpha}{\partial Q_i}
= \sum_{k=1}^{3N_{\mathrm{at}}}
   \frac{\partial\mu_\alpha}{\partial x_k}\,
   \frac{L_{ki}}{\sqrt{m_a}},
\label{eq:d1_trans}
\end{equation}

where $L_{ki}$ are the mass-weighted Hessian eigenvectors and
$a = \lceil k/3\rceil$ is the atom index. The result is a
$(3N_{\mathrm{at}} \times 3)$ matrix; only the $M$ vibrational
rows are retained.

The second derivatives are supplied in a mixed representation:
$\partial^2\mu_\alpha/\partial Q_i\,\partial x_k$, with the first index
already in normal coordinates. Only the second Cartesian index requires
transformation:

\begin{equation}
\frac{\partial^2\mu_\alpha}{\partial Q_i\partial Q_j}
= \sum_{k=1}^{3N_{\mathrm{at}}}
   \frac{\partial^2\mu_\alpha}{\partial Q_i\partial x_k}\,
   \frac{L_{kj}}{\sqrt{m_a}} .
\label{eq:d2_trans}
\end{equation}

The ORCA-provided array has dimensions $M \times 3N_{\mathrm{at}} \times 3$.
Internally, the resulting $M \times M \times 3$ array is divided by 2 to
absorb the factor of $1/2$ from the Taylor expansion, so that the VCI dipole
routine multiplies directly by the stored value.

\subsubsection{Dipole Matrix Elements in the VCI Basis}
\label{sec:vci_dipole_elements}

The full dipole matrix element between configurations $\bmv$ and $\bn$
consists of three contributions:

\paragraph{Linear dipole (one-body):}
\begin{equation}
D_\alpha^{(1)}(\bmv,\bn) =
  \sum_{i=1}^{M} \mu_\alpha^{(i)}\;
  \MI{i}(m_i,n_i;1)\;\Sexcl{i}.
\label{eq:D1}
\end{equation}

\paragraph{Diagonal second derivative (one-body):}
\begin{equation}
D_\alpha^{(2a)}(\bmv,\bn) =
  \sum_{i=1}^{M} \mu_\alpha^{(ii)}\;
  \MI{i}(m_i,n_i;2)\;\Sexcl{i}.
\label{eq:D2a}
\end{equation}

\paragraph{Off-diagonal second derivative (two-body):}
\begin{equation}
D_\alpha^{(2b)}(\bmv,\bn) =
  \sum_{i<j} 2\,\mu_\alpha^{(ij)}\;
  \MI{i}(m_i,n_i;1)\;\MI{j}(m_j,n_j;1)\;
  \Sexcl{\{i,j\}} .
\label{eq:D2b}
\end{equation}

At the HO and VSCF levels, only $D_\alpha^{(1)}$ is used, \textit{i.e.}, the linear dipole
surface. At the VCI, S-VCI, and SA-VCI levels, all three contributions
are evaluated. The quadratic terms $D_\alpha^{(2a)}$ and $D_\alpha^{(2b)}$ are
essential for overtone and combination band intensities, which are strictly zero
in the double-harmonic approximation, and for correctly redistributing oscillator
strength among Fermi-resonant states.

The configuration-space dipole matrix is transformed to the eigenstate basis using
BLAS: first $\mathbf{t}_\alpha = \boldsymbol{\mu}_\alpha \cdot \mathbf{c}_0$
(DGEMV), then $\langle\Psi_I|\hat\mu_\alpha|\Psi_0\rangle
= \mathbf{c}_I^T \cdot \mathbf{t}_\alpha$ (DDOT). Intensities are globally normalised by default for convenience in spectral visualisation and peak assignment. The normalisation is performed according to:

\begin{equation}
I_{0I}^{\mathrm{norm}} = 100 \, \tilde{I}_{0I} \Bigg/ \sqrt{\sum_{\Delta E \leq E_{\mathrm{cutoff}}} \tilde{I}_{0J}^2},
\end{equation}

where $\tilde{I}_{0I} = \sum_{\alpha} |\mu_{\alpha,0I}|^2$, $\Delta E = E_I - E_0$ is the transition energy from the ground state, and $E_{\mathrm{cutoff}}$ is a user-specified energy threshold with a default value of 4500 cm$^{-1}$. This ensures that the relative intensities are independent of higher-energy states that are often irrelevant for practical comparisons. Transitions with energy above $E_{\mathrm{cutoff}}$ are not displayed in the spectrum. The normalisation is explicitly defined so that the ratios between peaks are invariant with respect to the total number of calculated states. Absolute intensities, e.g. in units of km mol$^{-1}$ can be obtained by modifying the source code to bypass this normalisation; as \vibra{} is open-source, such modifications are straightforward for users requiring absolute values.
\section{Program Architecture and User Utilities}
\label{sec:utilities}

\vibra{} consists of a Fortran 90/95 computational engine and a Python~3
graphical interface. The computational engine is organized into modular
components, each responsible for a distinct stage of the anharmonic
vibrational spectroscopy workflow. The principal modules are summarized in
Table~\ref{tab:program_architecture}.

\begin{table}[htbp]
\centering
\small
\caption{Main computational modules in the \vibra{} Fortran engine.}
\label{tab:program_architecture}
\begin{tabularx}{\linewidth}{@{}lY@{}}
\toprule
Module & Role \\
\midrule
\code{main.f90}
& Entry point and orchestration layer for the execution workflow. \\

\code{read_input.f90}
& Keyword-based input parser with validation of user-specified options. \\

\code{read_orca.f90}
& Parser for ORCA \code{.vpt2} output files. This module performs Hessian
diagonalization, handles dipole transformations, and writes
\code{normal_mode.txt}. \\

\code{get_combination.f90}
& Configuration-enumeration routines, including recursive stars-and-bars
combinatorial counting, unique-element decomposition, and degeneracy-factor
evaluation. \\

\code{integrals.f90}
& Analytical harmonic-oscillator matrix elements used in the construction of
the vibrational Hamiltonian. \\

\code{one_mode_operation.f90}
& VSCF mean-field routines, including OpenMP-parallelized one-mode operations. \\

\code{vci.f90}
& Central VCI kernel. This module contains the shared
\code{compute_H_element} routine and supports full VCI, selected VCI with
EN-PT2 screening, symmetry-adapted VCI, and dipole-property calculations. \\

\code{jacobi.f90}
& LAPACK-based eigensolver wrappers, including \code{DSYEVD} and
\code{DSYEVR}, with automatic workspace queries and eigenvector normalization. \\

\code{symmetry.f90}
& Symmetry-handling routines, including point-group character tables, symmetry
operations, direct-product tables, normal-mode irrep assignment by
principal-axis-frame projection, and block-diagonal VCI dispatch. \\
\bottomrule
\end{tabularx}
\end{table}

The engine produces three output files, which can be utilized by the visualisation interface:

\begin{itemize}
  \item \texttt{vscf.out} --- complete log: VSCF convergence histories, modal
    coefficient matrices, VCI configuration list, eigenstate energies, and
    the three leading CI coefficients with quantum-number assignments;
  \item \texttt{intensities.txt} --- transition frequencies (cm$^{-1}$) and
    normalised intensities for HO, VSCF, and VCI/SCI methods;
  \item \texttt{normal\_mode.txt} --- equilibrium geometry (\AA) and Cartesian
    displacement vectors for all $3N_{\mathrm{at}}$ modes.
\end{itemize}

\subsection{Graphical User Interface}
\label{sec:GUI}

\begin{figure}[h!]
    \centering
    \fbox{\includegraphics[width=0.75\linewidth]{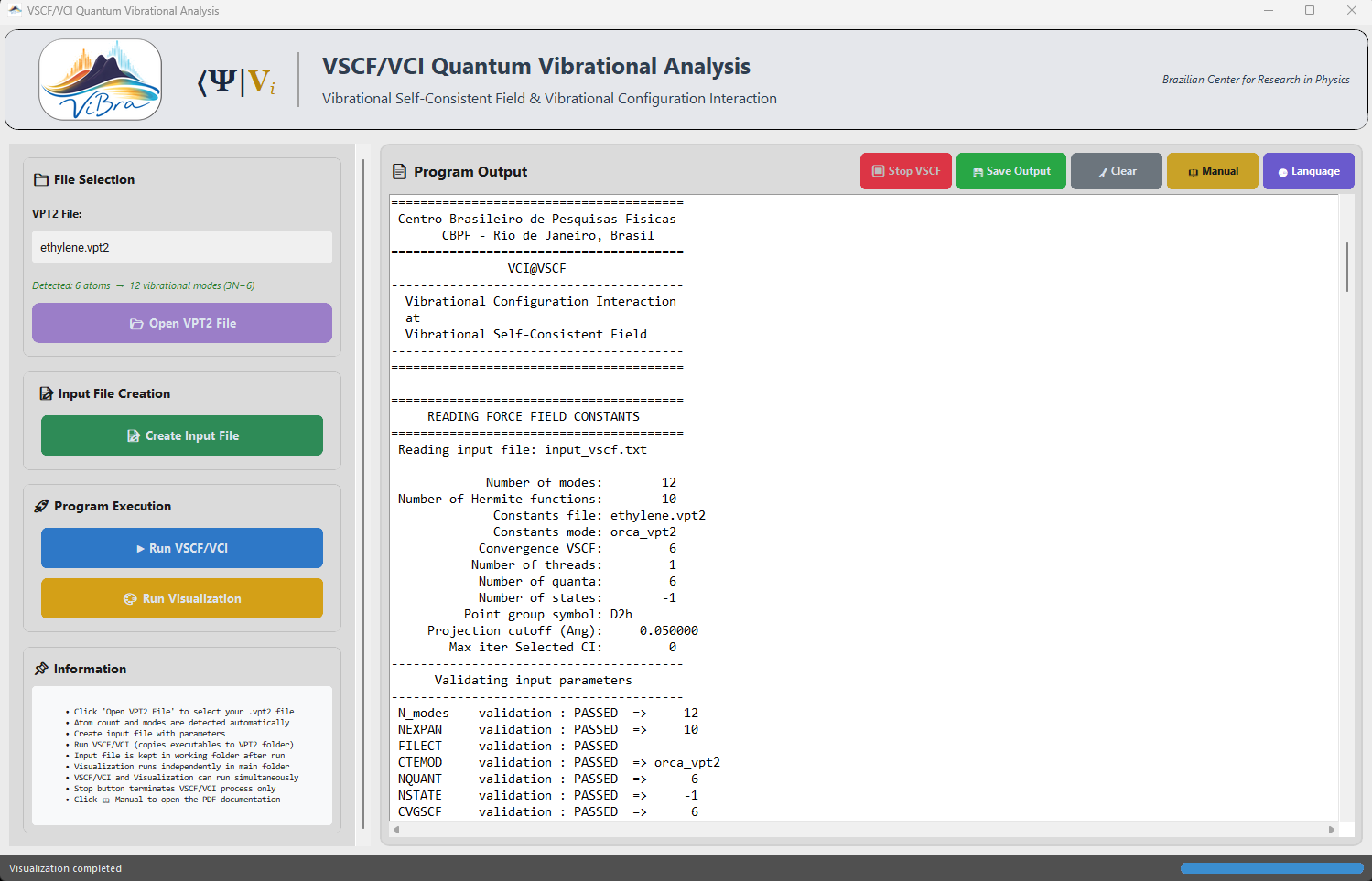}}
    \caption{Graphical interface of \vibra{}, showing the capability to directly load \texttt{ORCA} VPT2 output files, generate input files, and execute the code.}
    \label{figa}
\end{figure}

A CustomTkinter application, Fig. \ref{figa} (\texttt{VSCFVCIApp}) simplifies job preparation.
The user navigates to the ORCA \texttt{.vpt2} file; the working directory is set
automatically. A modal dialog collects all keywords (Table~\ref{tab:keywords}),
validates their ranges with tooltip guidance, and writes \texttt{input\_vscf.txt}.
The Fortran executable is launched as a subprocess in a background thread, with
real-time output streaming to a scrollable panel. A stop button terminates a
running job, and the panel supports save and clear operations.

\begin{table}[H]
\centering\small
\caption{Principal input keywords of \vibra.}
\label{tab:keywords}
\begin{tabularx}{\textwidth}{l l X}
\toprule
\textbf{Keyword} & \textbf{Type} & \textbf{Description}\\
\midrule
\texttt{NMODES} & int   & Number of vibrational modes ($M$).\\
\texttt{NEXPAN} & int   & HO basis size per mode ($N_{\mathrm{exp}}$).\\
\texttt{FILECT} & str   & Path to ORCA \texttt{.vpt2} file.\\
\texttt{CTEMOD} & str   & Format (\texttt{orca\_vpt2}).\\
\texttt{NQUANT} & int   & Maximum total quanta ($N_q$); $\le 0$ disables VCI.\\
\texttt{NSTATE} & int   & Number of eigenstates ($\le 0$ = all).\\
\texttt{CVGSCF} & int   & VSCF convergence exponent ($10^{-C}$ cm$^{-1}$).\\
\texttt{THREAD} & int   & Number of OpenMP threads.\\
\texttt{PGROUP} & str   & Point group (C1, Cs, Ci, C2, C2h, C2v, D2, D2h).\\
\texttt{PROJCT} & real  & Projection cutoff for symmetry detection (\AA).\\
\texttt{MAXSCI} & int   & $N_{\mathrm{sel}}$ for S-VCI. Default: 100.\\
              &       & Usage: \texttt{MAXSCI [N] [mode] [ref]}\\
              &       & \texttt{N > 0, mode = auto}: iterative selection with CI reference (\texttt{ref} = \texttt{s}, \texttt{d}, \texttt{t}, \texttt{q}).\\
              &       & \texttt{N > 0, mode = list}: iterative selection with list reference, no \texttt{ref} needed. \\
              &       & \texttt{N = 0}: full VCI (no S-VCI),  no \texttt{mode} and no \texttt{ref} needed.\\
              &       & \texttt{N = 0, mode = list}: full VCI using user-provided state list,  no \texttt{ref} needed.\\
\bottomrule
\end{tabularx}
\end{table}

\subsection{Interactive Spectral Viewer}
\label{sec:viewer}

The viewer, Fig. \ref{figb} (\texttt{MainApplication}, Python 3, Matplotlib, 3Dmol.js) reads the
three output files and provides a rich environment for analysis and comparison:

\begin{figure}[h!]
    \centering
    \fbox{\includegraphics[width=0.75\linewidth]{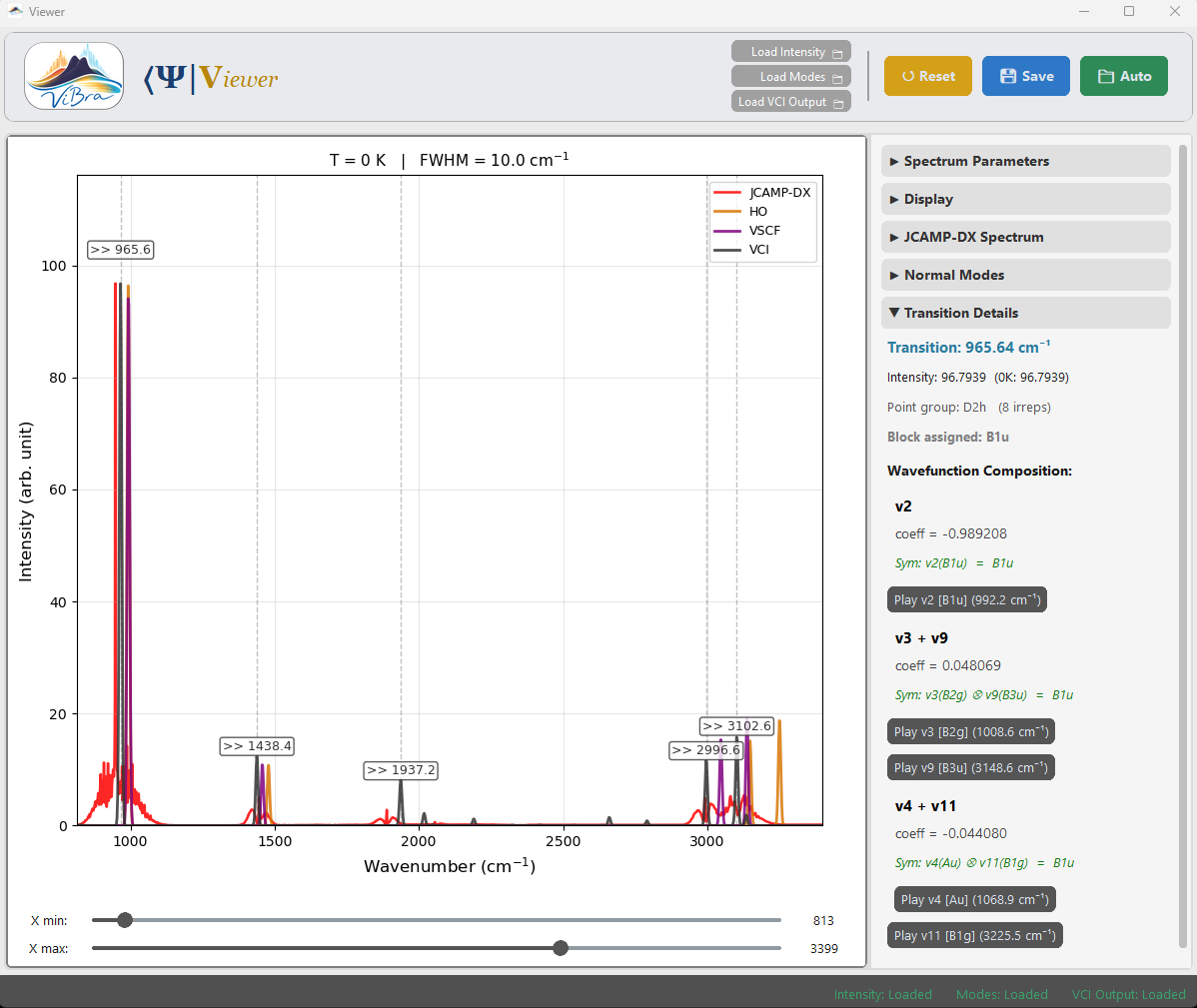}}
    \caption{Dedicated \vibra{} spectrum visualizer, highlighting the capability to directly compare HO, VSCF, and VCI calculations with experimental data, while seamlessly analyzing the peak composition associated with each assigned mode. The displayed results correspond to the ethylene molecule, where the full VCI method was applied including up to 6 quanta, and 10 HO functions per mode were used in the VSCF calculation. Experimental data were retrieved from the National Institute of Standards and Technology database \cite{coblentz_ethylene_1964}.}
    \label{figb}
\end{figure}

\begin{itemize}
  \item \textbf{Gaussian broadening:} each transition at frequency $\nu_k$ with
    intensity $I_k$ is represented as
    $S(\nu) = \sum_k I_k \exp[-(\nu-\nu_k)^2/(2\sigma^2)]$,
    with $\sigma = \mathrm{FWHM}/(2\sqrt{2\ln 2})$. FWHM is adjustable via stepper
    and direct text entry.
  \item \textbf{Temperature-dependent spectra:} for $T > 0$ K, VCI intensities are
    Boltzmann-weighted: $I_k^{\mathrm{eff}} = I_k\,(p_0 - p_k)$, where
    $p_k = \exp(-h c\,\nu_k/k_B T)/Z$ and $hc/k_B = 1.438777$ cm$\cdot$K.
    Temperature is adjustable from 0 to 950 K.
  \item \textbf{JCAMP-DX overlay:} experimental spectra in JCAMP-DX format
    (\texttt{.jdx}/\texttt{.dx}) can be loaded. Transmittance is automatically
    converted to absorbance, negative values are clipped, and the result is
    normalised.
  \item \textbf{ALS baseline correction:} for experimental spectra exhibiting
    a sloping baseline, an Asymmetric Least Squares (ALS) algorithm~\cite{eilers2003}
    estimates and subtracts a smooth baseline. The cost function
    $Q(\mathbf{z}) = \sum_i w_i(y_i - z_i)^2 + \lambda\|\mathbf{D}\mathbf{z}\|^2$
    is minimised iteratively, where $\mathbf{D}$ is the second-difference matrix.
    The smoothness parameter $\lambda$ and asymmetry parameter $p$ are adjustable
    via steppers with real-time redisplay.
  \item \textbf{Peak inspection:} clicking a VCI peak label displays the frequency,
    intensity, and three leading CI coefficients with quantum-number assignments
    (e.g., ``$v_1 + 2v_3$''). Buttons launch 3D normal-mode animations for any
    involved mode.

  \item \textbf{3D normal-mode animation:} sinusoidal displacement trajectories
    (60 frames, adjustable amplitude) are generated from \texttt{normal\_mode.txt},
    saved as JSON, and rendered via an auto-generated HTML page with embedded
    3Dmol.js viewer supporting play/pause, frame slider, speed control, and
    adjustable rendering styles and atomic radii. See Fig. \ref{figc}.

        \begin{figure}[h!]
            \centering
            \fbox{\includegraphics[width=0.5\linewidth]{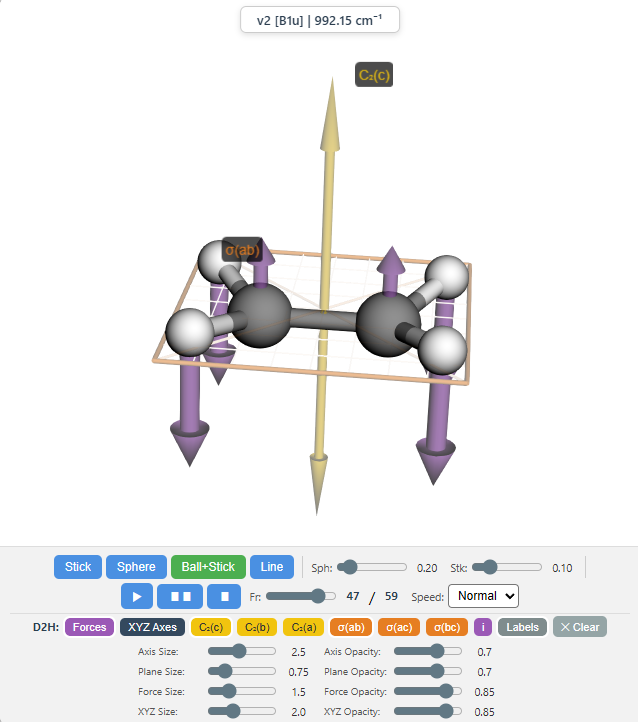}}
            \caption{Normal mode animator, featuring the ability to display force vectors and symmetry elements whenever a point group other than C1 is employed.}
            \label{figc}
        \end{figure}

  \item \textbf{Data export:} the displayed spectrum can be saved as a text file
    and as a 300 dpi PNG image.
\end{itemize}

\section{Computational Optimisation Strategies}
\label{sec:optim}

\vibra{} employs several targeted algorithmic optimisations that are essential
for making VCI calculations on molecules with $M \approx 30$--$40$ and larger tractable.
The complete rationale for each strategy is given in the manual, provided as supporting information;
the most impactful are summarised here.

\textbf{Overlap Cache in VSCF.}
At the beginning of each VSCF iteration, all scalar overlaps
$\langle\varphi_{n_j}^{(j)}|Q_j^c|\varphi_{n_j}^{(j)}\rangle$
($c=0,\dots,4$) are precomputed and cached. The $X_i^{(p)}$ coefficients are then
assembled via $O(1)$ look-ups rather than repeated $O(N_{\mathrm{exp}}^2)$ double sums.

\textbf{Vectorised potentials and nonzero flags.}
Force constants $V_{ijk}$ and $V_{ijkl}$ are permutationally symmetric, so a
dense $M^3$ or $M^4$ array wastes memory and access time on redundant
duplicates. Instead, each is flattened into a 1D array of length $T_3$
(respectively $T_4$), containing one entry per sorted index combination
$i\le j\le k$ (respectively $i\le j\le k\le l$), together with precomputed
metadata: the distinct mode indices appearing in the term, their
multiplicities (e.g.\ $V_{iij}$ has mode $i$ with multiplicity 2 and mode $j$
with multiplicity 1), and a boolean flag marking whether the term exceeds a
numerical threshold and is therefore treated as
nonzero. This flattening removes redundant storage and lets the
Hamiltonian-construction loop iterate directly over the $T_3+T_4$ distinct
terms rather than the full $M^3+M^4$ index space, with the nonzero flag
skipping negligible terms at no extra cost.

\textbf{Sparse pair list.}
Before building $H$, the code makes a single pass over all configuration
pairs and records only those with $d \le 4$ differing modes, storing for each
such pair the indices of the modes that differ. The Hamiltonian-construction
loop then iterates directly over this precomputed list rather than over all
$|\mathcal{S}|(|\mathcal{S}|+1)/2$ pairs, so pairs that are known in advance
to give zero coupling are never visited.

\textbf{Inverted potential index.}
For a pair differing in $d$ modes, a cubic or quartic term contributes only
if it involves \emph{all} $d$ differing modes; otherwise modal orthogonality
forces at least one factor to vanish. Scanning the full list of $T_3+T_4$
potential terms for every pair would waste most of the work rejecting
non-matching terms. Instead, each mode is pre-indexed with the list of
potential terms it appears in. For a given pair, only the terms attached to
the \emph{first} differing mode are examined, and each is quickly checked for
the remaining $d-1$ differing modes. This reduces the average per-pair
potential scan from $O(T_3+T_4)$ to $O\!\left(\frac{T_3+T_4}{M^{\,d-1}}\right)$ since a term touches a given mode with probability $\sim 1/M$.

\textbf{Prefix--Suffix Overlap Products.}
The spectator product $\Sexcl{S}$ (Eq.~\ref{eq:Sexcl}) is needed for every
Hamiltonian and dipole contribution. Prefix and suffix arrays of
$o_k = \MI{k}(m_k,n_k;0)$ are computed once per matrix element in $O(M)$ time,
enabling $O(1)$ evaluation of any excluded-mode product. For multi-mode exclusion,
the total product is divided by the excluded factors; a fallback explicit product
handles the case where any overlap vanishes.

\textbf{OpenMP Parallelism.}
The Hamiltonian construction loop over the sparse pair list, the EN-PT2 screening
in SCI, and the dipole matrix construction are parallelised with dynamic scheduling
and empirically tuned chunk sizes (16--256). Dynamic scheduling is essential because
pairs with $n_{\mathrm{diff}}=0$ (diagonal) evaluate all nonzero potential terms
while high-$n_{\mathrm{diff}}$ pairs evaluate far fewer, creating significant load
imbalance. The VSCF stage runs in serial since, with the overlap cache, it completes
significantly faster.

\textbf{LAPACK/BLAS Integration.}
The VSCF modal matrices are diagonalised with \texttt{DSYEVD}. For VCI/S-VCI,
\texttt{DSYEVD} is used when all eigenvalues are requested; \texttt{DSYEVR} computes
only the lowest $N_{\mathrm{state}}$ eigenvalues (if point group is C1 and no external list is given) for partial spectra, saving both
memory and time. Dipole transformations use \texttt{DGEMV} and \texttt{DDOT}.

\textbf{Memory Considerations.}
\label{sec:memory}
The full VCI Hamiltonian is stored as a dense symmetric matrix,
which limits the practical configuration space size to
$|\mathcal{S}| \sim 10^4$--$10^5$ depending on available memory
(the matrix requires $|\mathcal{S}|^2 \times 8$~bytes).  The SCI
approach directly alleviates this by reducing the matrix dimension.
The modal integral table is compact
($M \times (N_q{+}1)^2 \times 6$ entries).  The sparse pair list
and inverted index use integer arrays whose size is data-dependent
but typically modest compared to the Hamiltonian itself.
\section{Validating and using \vibra{}}

\subsection{Validation of the \vibra{} implementation}

We carried out a systematic validation of the \vibra{} code through four complementary benchmarks, progressing from the simplest VSCF level to fully correlated VCI and symmetry-adapted formulations. All geometry files, \textsc{ORCA} inputs and outputs, \vibra{} inputs and outputs, as well as the \textsc{Crystal23} output for the water, are provided as Supporting Information for numerical comparison. For the water benchmark (step 1 and 2), the agreement with reference data is within numerical precision; therefore, we do not report the numerical outputs in the main text, and the reader is referred to the SI for full details. For the validation of the SA-VCI and S-VCI implementations (step 3 and 4), the complete dataset, comprising all vibrational modes, VCI coefficients, and energy contributions for each molecule is extensive and is provided in its entirety in the Supporting Information and in a public Zenodo dataset\cite{QuantumVibData}. In the following section \ref{usevibra}, we present the ethylene molecule as a representative case study to illustrate the key findings.

\subsubsection*{Step 1: Validation of the VSCF implementation}

\textbf{Goal:} to verify the correctness of the VSCF module in \vibra{}. \textbf{Action:} we first validated the VCI@VSCF implementation by benchmarking VSCF ground-state energies of a water molecule computed with \textsc{Crystal23} at the B3LYP/6-31G* level of theory. VSCF calculations were performed using a basis set of 10 harmonic oscillator (HO) functions per vibrational mode. We obtained the reference data (i.e. cubic and quartic force constants and normal mode frequencies) from the \textsc{Crystal23} website, where they are provided as part of the anharmonic vibrational tutorial \cite{erba2019, erba_mitoli_anharmonic_tutorial}. Therefore, we performed no computations at this level of theory, nor did we make use of \textsc{Crystal23}. \textbf{Comparison:} we compared ground-state energies, fundamental transition energies, and optimized modal coefficients. The computed energies show excellent agreement with the reference values, with deviations in the range of 0.01–0.001 cm$^{-1}$. The modal coefficients differ on the order of $10^{-3}$ or lower, also indicating excellent agreement. \textbf{Conclusion:} the VSCF implementation in \vibra{} is correct and reproduces reference data with numerical accuracy.

\subsubsection*{Step 2: Validation of the VCI implementation}

\textbf{Goal:} to verify the correctness of the VCI diagonalization and eigenstate analysis in \vibra{}. \textbf{Action:} we analyzed VCI energies and eigenstate coefficients using the same water molecule benchmark. A total excitation truncation of up to 5 quanta was utilized. \textbf{Comparison:} the observed discrepancies between \vibra{} and \textsc{Crystal23} are of the same magnitude as those obtained at the VSCF level and are attributed to numerical noise. The most important configurations, together with their numerical contributions, for each VCI state were correctly identified. \textbf{Conclusion:} These results demonstrate that the entire workflow, from the initial VSCF step to the final diagonalization of the VCI Hamiltonian matrix, is correctly implemented in \vibra{}. In particular, the computed eigenvalues (energies) and eigenvectors (VCI coefficients) are in full agreement with reference results, confirming the correctness of the implementation.

\subsubsection*{Step 3: Validation of the SA-VCI implementation}

\textbf{Goal:} To validate the symmetry-adapted VCI (SA-VCI) formulation against full VCI calculations. \textbf{Methodology:} We employed semi-quartic force fields, defined here as cubic force constants together with quartic contributions in which at least two vibrational modes are identical. Geometry optimizations, energy minimizations, and force constant computations were carried out using \textsc{ORCA}~6.1 at the wB97X-D4/aug-cc-pVTZ level of theory. The test set comprised molecules spanning several point-group symmetries: HFCO ($C_s$), N$_2$H$_4$ ($C_2$), H$_2$O ($C_{2v}$), \textit{trans}-N$_2$H$_2$ ($C_{2h}$), and C$_2$H$_4$ ($D_{2h}$). \textbf{Comparison:} For all molecules examined, the low lying state energies obtained from full VCI and SA-VCI agree to within numerical noise, on the order of $10^{-2}\,\mathrm{cm}^{-1}$ or better. \textbf{Conclusion:} The SA-VCI formulation is correctly implemented and reproduces full VCI results to numerical precision. \textbf{Computational details:} Our choice of functional and basis set was driven not by the pursuit of highly accurate vibrational frequencies, but by considerations of balanced computational cost and methodological robustness. The range separated hybrid functional wB97X-D4 combined with the aug-cc-pVTZ basis set is widely used in benchmark vibrational studies, including several entries at the CCCBDB/NIST frequency database~\cite{nist_cccbdb_2022}. In the VSCF step, we adopted a basis of 10 harmonic oscillator functions per vibrational mode, a choice motivated by the rapid convergence of VSCF energies with respect to the number of HO functions and by the fact that this basis provides a balanced description of vibrational states with few quanta of excitation per mode~\cite{erba2019}. For both the full VCI and SA-VCI expansions, we included configurations with up to 4 quanta. \textbf{Important note on symmetry handling:} Despite the tight convergence criteria for minimizing the geometries, no symmetry constraints were imposed during the \textsc{ORCA} computations; geometries, Hessians, and force constants were evaluated entirely without point-group restrictions. This choice was deliberate for the present validation study: our objective is also to probe the robustness of the SA-VCI implementation against minute numerical asymmetries that may arise in the quartic force field, we caution that for production level applications users are strongly advised to enforce appropriate symmetry constraints during the electronic structure calculations. Even tiny deviations introduced by a non symmetrized quartic force field can accumulate, leading to larger errors, specially for higher energy states.

\subsubsection*{Step 4: Three-tier benchmark of intensities and variational selection}

\textbf{Goal:} To validate the intensity calculations and the selected-state VCI (S-VCI) procedure. \textbf{Action:} We carried out a three tier benchmark that begins from the standard double harmonic approximation and progressively includes higher order effects. The electronic structure calculations were the same as those described in Step~3, and all molecules in the test set were benchmarked. \textbf{Tier 1 (DHA validation):} We benchmarked intensities obtained with \textsc{ORCA}~6.1 and \vibra{} within the harmonic approximation using linear dipole derivatives, finding identical results for all molecules, noting that \vibra{} intensities are internally normalized. This step corresponds to the full double harmonic approximation (harmonic frequencies + linear dipole derivatives). \textbf{Tier 2 (Electrical anharmonicity):} We analyzed \vibra{} energies and intensities including also second-order dipole derivatives, both within the full VCI framework and within the symmetry-adapted VCI (SA-VCI) formulation. For all molecules, the resulting energies agree within numerical noise, on the order of $10^{-2}\,\mathrm{cm}^{-1}$ or lower. The intensities obtained from SA-VCI and full VCI are identical within numerical precision. \textbf{Tier 3 (S-VCI variational validation):} To validate the S-VCI (selected states via EN-PT2), we analyzed the variational behavior of the method by systematically increasing the number of included states. As expected for a variational approach, all state energies decreased monotonically upon enlarging the configuration space by increasing \texttt{MAXSCI}. \textbf{Conclusion:} The intensity implementations (both DHA and with second-order derivatives) are correct, and the S-VCI selection procedure behaves variationally as expected.

\subsection{Using \vibra{}}
\label{usevibra}

To demonstrate the efficiency of the software, particularly for VCI and its variants, we consider the ethylene (C$_2$H$_4$) molecule. Its $D_{2h}$ symmetry provides a clear example of the computational advantages gained from block diagonalization. Since VSCF calculations are computationally inexpensive, we do not provide a detailed performance analysis for this step. We employ a fixed basis of 10 harmonic oscillator functions per vibrational mode and an energy convergence threshold of $10^{-6}\,\mathrm{cm}^{-1}$ in the VSCF procedure preceding all VCI and VCI-variant calculations. It is important to note that all frequency calculations performed with \vibra{} were carried out on an 11th Gen Intel\textregistered{} Core\texttrademark{} i9-11900K processor, with the CPU frequency capped at 3.5 GHz. The machine was equipped with 2x32 GB of RAM running at 3200 MHz, and cache sizes of 640 KB (L1), 4.0 MB (L2), and 16.0 MB (L3). The code is compiled with \texttt{ifx}, Intel's next-generation LLVM-based Fortran compiler, using \texttt{-O3} for aggressive optimisation including loop transformations and vectorisation. Parallelism is enabled at multiple levels: \texttt{/Qopenmp} activates explicit OpenMP directives, \texttt{/threads} enables compiler auto-parallelisation of loops, and \texttt{/Qmkl:parallel} links the threaded Intel Math Kernel Library for parallel BLAS/LAPACK routines. The \texttt{/Qm64} flag targets the 64-bit instruction set, while \texttt{/heap-arrays} allocates temporary arrays on the heap to prevent stack overflow from large automatic arrays. Finally, \texttt{/fpscomp:logicals} enforces compatibility with older Intel Fortran logical behaviour. A pre-compiled executable is provided alongside the source code in the repository.

Using this computational setup, we examine the dependence of basis size, energy convergence, and diagonalization cost on the maximum number of quanta included in the VCI expansion, as summarized in Table~\ref{AA}. Note that a single core was set for this benchmark. The Hamiltonian dimension increases rapidly as the maximum excitation level is raised. Increasing the maximum number of allowed quanta from 4 to 5, and subsequently
from 5 to 6, approximately triples the number of vibrational configurations
at each stage, resulting in an approximately ninefold increase in the
number of Hamiltonian matrix elements (and hence memory footprint) for
every additional maximum quanta.

The zero point energy (ZPE) and the first excited state energy ($E_1$) both decrease monotonically as the maximum excitation level is increased, as expected from the systematic enlargement of the VCI configuration space. However, their convergence rates differ significantly. The ZPE is already effectively converged when a maximum of 3 quanta is included, whereas $E_1$ requires at least 4 quanta to reach a practically stable value. This behavior is consistent with the semi quartic character of the force field. Because the reductions in the ZPE and $E_1$ are not proportional, the transition energy, $\Delta E = E_1 - \mathrm{ZPE}$, does not exhibit monotonic convergence. Instead, it increases by approximately 60 cm$^{-1}$ when the maximum excitation level is increased from 2 to 3 quanta before decreasing to a value below that obtained in the calculation with 2 quanta.

The apparent stabilization at 4 quanta should not be interpreted as strict numerical convergence. Rather, it indicates that the remaining changes are sufficiently small for the practical applications considered here and do not justify the substantial additional computational expense associated with including higher excitation levels. Therefore, a maximum of 4 quanta provides an appropriate compromise between accuracy and computational efficiency for routine calculations. When higher numerical accuracy is required, however, calculations including 5 or 6 quanta may still be necessary.

This need for increased accuracy must be balanced against the rapidly increasing computational cost. Both the memory requirements and the CPU time increase dramatically as the maximum number of allowed quanta is raised from 4 to 6, with matrix diagonalization becoming the dominant computational bottleneck. For the largest calculation, corresponding to a maximum of 6 quanta and a Hamiltonian containing 18564 vibrational configurations, diagonalization with \texttt{DSYEVD} required 1400 s (23.3 min) out of a total runtime of 1438 s.

To reduce this computational expense, a partial diagonalization was also performed using \texttt{DSYEVR}, requesting only the lowest 20 eigenvalues. This approach avoids constructing and storing the complete set of eigenvectors, reducing the diagonalization time to 632 s (10.5 min) and the total runtime to 670 s, as reported in the final column of Table~\ref{AA}. Although the eigenvectors are no longer available for subsequent intensity calculations, the reduction in computational cost is substantial and becomes increasingly advantageous when only a limited number of state energies is required. Consequently, for applications that require only the ZPE, such as the calculation of zero point energy corrections to electronic energies in chemical reaction transition barrier calculations, partial diagonalization provides a particularly efficient strategy.


\begin{table}[htbp]
\centering
\small
\setlength{\tabcolsep}{4pt}
\caption{
Convergence of VCI energies and computational cost with increasing excitation
quanta for ethylene. The final row corresponds to partial diagonalization of
the 6-quanta Hamiltonian requesting only the lowest 20 eigenvalues. All
energies are reported in cm$^{-1}$ and timings in seconds.
}
\label{tab:ethylene_quanta_convergence}
\begin{tabular}{lrrrrrrr}
\toprule
\# Quanta
& \# States
& $t_{\mathrm{diag}}$
& $t_{\mathrm{tot}}$
& ZPE
& $E_1$
& $\Delta E$ \\
\midrule
2 & 91    & 0    & 0    & 11191.4056 & 12036.0327 & 844.6271 \\
3 & 455   & 0    & 0    & 11129.7226 & 12034.6711 & 904.9484 \\
4 & 1820  & 1    & 2    & 11127.0028 & 11959.2915 & 832.2886 \\
5 & 6188  & 38   & 44   & 11126.7737 & 11953.6805 & 826.9069 \\
6 & 18564 & 1400 & 1438 & 11125.5159 & 11952.9510 & 827.4351 \\
6(20) & 18564 & 632  & 670  & 11125.5159 & 11952.9510 & 827.4351 \\
\bottomrule
\end{tabular}
\label{AA}
\end{table}

The full diagonalization of the 6-quanta Hamiltonian, although manageable for ethylene, remains computationally demanding. This motivates the use of selected VCI (S-VCI) and symmetry-adapted VCI (SA-VCI) to further reduce the cost while preserving accuracy. Table~\ref{tab:data2} compares the performance of both approaches using 8 OpenMP threads; the reported real time refers to the total wall-clock time encompassing state selection (S-VCI) or symmetrization (SA-VCI), Hamiltonian construction, diagonalization, and intensity evaluation. In the S-VCI columns, the number following the slash denotes the value of the \texttt{MAXSCI} keyword, which controls how many states are retained by the perturbative selection criterion. Increasing \texttt{MAXSCI} systematically enlarges the selected space and drives the energies toward the full-VCI limit. Already at \texttt{MAXSCI}=50, the ZPE and $E_1$ lie within 1.6 and 2.3~cm$^{-1}$ of the full-VCI result, respectively, and further increasing the keyword from 100 to 300 yields only marginal improvement. This indicates that low \texttt{MAXSCI} values may filter too aggressively, discarding states that contribute significantly to the correlation energy (see Section~\ref{validation_qpu_h20}); a moderate threshold (here 50) already captures the essential physics. The optimal value is molecule-dependent and should be calibrated, when possible, by comparing a small-basis S-VCI calculation against full VCI, as done here. Notably, within the S-VCI kernel, we used the default VCISD reference, with external states generated up to \texttt{NQUANT}=6.

\begin{table}[htbp]
\centering
\small
\caption{Performance comparison of selected VCI (S-VCI) and symmetry-adapted VCI (SA-VCI) against full VCI for ethylene at the 6-quanta level. S-VCI rows are labelled by the \texttt{MAXSCI} keyword value. All calculations used 8 OpenMP threads.}
\label{tab:data2}
\begin{tabular}{lrrrrr}
\hline
Method & \# States & CPU time (s) & Real time (s) & ZPE (cm$^{-1}$) & $E_1$ (cm$^{-1}$) \\
\hline
S-VCI/50        & 3193  & 31   & 4   & 11127.0820 & 11955.2475 \\
S-VCI/100       & 5921  & 138  & 18  & 11126.7083 & 11953.8296 \\
S-VCI/150       & 7895  & 208  & 28  & 11126.5374 & 11953.6018 \\
S-VCI/200       & 9241  & 303  & 40  & 11126.5056 & 11953.4682 \\
S-VCI/250       & 10339 & 414  & 55  & 11126.4265 & 11953.4365 \\
S-VCI/300       & 11140 & 494  & 66  & 11126.3749 & 11953.4153 \\
Full VCI        & 18564 & 2005 & 265 & 11125.5159 & 11952.9510 \\
SA-VCI ($D_{2h}$) & 18564 & 89  & 13  & 11125.5202 & 11952.9591 \\
SA-VCI ($C_i$)    & 18564 & 601 & 81  & 11125.5177 & 11952.9550 \\
\hline
\end{tabular}
\end{table}
Turning to symmetry-adapted VCI, the $D_{2h}$ point group of ethylene partitions the Hamiltonian into eight irreducible representation blocks. In the ideal scenario where all blocks are of equal size, each block has dimension $N/8$, and the cubic scaling of diagonalization (e.g., using the DSYEVD subroutine) reduces the total cost by a factor of $8 \times (1/8)^3 = 1/64$, yielding a theoretical maximum speedup of $64\times$.
In our implementation, the symmetry blocks are processed sequentially, but each block diagonalization is executed using all available OpenMP threads, rather than distributing blocks across threads and subdividing the thread team. This design ensures that the full multi-threaded capability of the diagonalizer is applied to every block, avoiding load imbalance when the number of symmetry operations does not evenly divide the thread count. 

In practice, the observed reduction in real time is approximately $20\times$ (from 265~s for full VCI to 13~s for SA-VCI/$D_{2h}$). The primary reason for $20\times$ speed up is simple: not all operations scale cubically, so a 64$\times$ speedup is never expected, it will always be less. Moreover, it should be noted that SA-VCI and full VCI employ different computational kernels, and the timings reported in Table~\ref{tab:data2} reflect the total execution time from initialization to completion, not solely the diagonalization step. Yet, the irreducible blocks are not perfectly equal in size, so the sum of cubed block dimensions exceeds $8\times(N/8)^3$, reducing the theoretical gain. Also, the non-diagonalization overhead (symmetrization, integral transformation, etc.) accounts for a larger fraction of the total time in the SA-VCI run, further diluting the speedup. Finally, the per-block matrices (around 2300$\times$2300) may be too small to keep all 16 threads fully utilized, leading to higher parallel overhead than in the full VCI diagonalization of the much larger unblocked matrix. These effects are partially offset by improved cache efficiency when blocks or specific operations requires arrays that fit into lower cache levels, but the net result is a reduction from the ideal 64$\times$ to $\sim$20$\times$. 

Notably, the gap between ideal and observed speedup narrows as the number of symmetry blocks decreases. Using a smaller symmetry group, such as $C_i$, yields only two blocks, a theoretical speedup of $4\times$, and an observed reduction to 81~s---an efficiency of 82\% relative to the ideal, compared to 31\% for the $D_{2h}$ case. The larger block size in $C_i$ restores diagonalization as the dominant cost, and the  $\sim$2300$\times$2300 blocks are replaced by matrices roughly 16 times larger,  $\sim$9200$\times$9200, which make better use of the available threads. This trend indicates that our sequential-block parallelization strategy performs best when blocks are large enough to saturate the multi-threaded overhead.

Importantly, the energies obtained with SA-VCI/$D_{2h}$, SA-VCI/$C_i$, and full VCI are numerically equivalent to within a few hundredths of a cm$^{-1}$, confirming that symmetry adaptation introduces no approximation beyond block diagonalization. A word of caution is warranted: assigning normal modes to irreps requires the correct point group. Using a lower symmetry than the true molecular symmetry (e.g., $D_{2h}$ for benzene instead of $D_{6h}$) can lead to incorrect irrep assignments, particularly for degenerate modes. It is therefore advisable to run a low-quanta SA-VCI calculation and verify that the resulting energies match the corresponding full-VCI values before proceeding to larger bases.


\section{Selected VCI as a bridge to quantum simulation algorithms}
\label{sec:quantum_bridge}

The selected-VCI machinery in \vibra{} can also be used in a quantum-centric workflow for vibrational problems. In a conventional \vibra{} calculation, the initial and selected spaces are generated classically using the procedures described in Section~\ref{sec:SCI}. In the quantum-centric setting considered here, a quantum sampling routine instead provides an externally generated seed set of vibrational configurations. \vibra{} then
acts as the classical selected-VCI engine: it can construct and diagonalize the projected anharmonic Hamiltonian directly in that seed or optionally enlarge it through EN-PT2-guided configuration selection before evaluating energies, eigenvectors, transition moments, and spectra.

This division of labour is closely related to recent sample-based quantum
diagonalisation strategies in electronic-structure theory. In sample-based
quantum diagonalisation (SQD) and sample-based Krylov quantum diagonalisation
(SKQD), a quantum processor is used to generate basis-state samples, while the
final Hamiltonian projection, diagonalisation, and observable evaluation are
performed classically~\cite{Kanno_2026,RobledoMoreno2025SQD,Yu2025SKQD}. The same
principle can be applied to vibrational structure. Instead of electronic
occupation strings, the samples correspond to vibrational configurations
\[
\bn=(n_1,\ldots,n_M),
\]
where $n_i$ labels the occupied modal or harmonic-oscillator level of mode
$i$. The sampled configurations define a compact subspace
$\mathcal{S}_{\mathrm{Q}}$, and \vibra{} evaluates
\begin{equation}
H^{(\mathcal{S}_{\mathrm{Q}})}_{\alpha\beta}
=
\langle \Phi_{\alpha} | \Hvib | \Phi_{\beta} \rangle,
\qquad
\Phi_{\alpha},\Phi_{\beta}\in\mathcal{S}_{\mathrm{Q}},
\label{eq:qsampled_projected_ham}
\end{equation}
using the same modal-integral and Hamiltonian-construction kernels employed in the classical VCI and S-VCI calculations. This distinction is central: \vibra{} does not replace the quantum sampling step. The quantum-facing routine supplies the initial seed configurations, while \vibra{} converts that seed into a projected VCI calculation and, when requested, systematically enriches it through classical EN-PT2-guided
selection.

\begin{figure}[t]
\centering

\definecolor{vibdone}{HTML}{2E8B57}
\definecolor{vibdonebg}{HTML}{DCEEDC}
\definecolor{vibprogress}{HTML}{C47A00}
\definecolor{vibprogressbg}{HTML}{FFF2CC}

\resizebox{\textwidth}{!}{%
\begin{tikzpicture}[
  node distance=1.55cm and 1.45cm,
  box/.style={
    rectangle, rounded corners,
    draw=vibblue, fill=viblightblue!35,
    very thick, align=center,
    minimum width=3.2cm, minimum height=1.05cm,
    font=\small
  },
  greenbox/.style={
    rectangle, rounded corners,
    draw=vibdone, fill=vibdonebg,
    very thick, align=center,
    minimum width=3.2cm, minimum height=1.05cm,
    font=\small
  },
  orangebox/.style={
    rectangle, rounded corners,
    draw=vibprogress, fill=vibprogressbg,
    very thick, align=center,
    minimum width=3.2cm, minimum height=1.05cm,
    font=\small
  },
  arrow/.style={-{Latex[length=3mm]}, very thick, draw=vibblue},
  dashedarrow/.style={-{Latex[length=3mm]}, very thick, dashed, draw=vibblue}
]

\node[greenbox] (force) {Electronic-structure\\inputs and QFFs\\ORCA / Crystal};
\node[greenbox, right=of force] (vib) {Vibrational\\Hamiltonian\\VSCF / VCI / S-VCI};
\node[orangebox, right=of vib] (pauli) {Qubit mapping\\modals to Pauli\\Qiskit-compatible operators};
\node[orangebox, below=of pauli] (frag) {Fragmentation and\\Trotter blocks\\time evolution};
\node[orangebox, left=of frag] (sample) {Quantum sampling\\SQD / SKQD\\configuration subspace};
\node[greenbox, left=of sample] (diag) {\vibra{} backend.\\Projected VCI; optional\\EN-PT2 enlargements};
\node[box, below=of diag] (obs) {Spectroscopy\\peak positions, intensities,\\assignments};
\node[box, below=of sample] (bench) {Validation\\full VCI, experiment\\};

\draw[arrow] (force) -- (vib);
\draw[arrow] (vib) -- (pauli);
\draw[arrow] (pauli) -- (frag);
\draw[arrow] (frag) -- (sample);
\draw[arrow] (sample) -- (diag);
\draw[arrow] (diag) -- (obs);
\draw[arrow] (obs) -- (bench);

\end{tikzpicture}
}
\caption{
Quantum--classical workflow connecting \vibra{} (green) to quantum vibrational
spectroscopy algorithms. The quantum-facing part of the workflow (yellow) maps the
vibrational Hamiltonian to qubit operators and uses fragmentation and
Trotterized time evolution to generate samples or time-domain data. The sampled
configuration space is then passed back to \vibra{}, which constructs and
diagonalises the projected VCI Hamiltonian and evaluates spectroscopic
observables (blue). These can then be validated against full VCI or even experimental data when available.
}
\label{fig:quantum_vibra_workflow}
\end{figure}
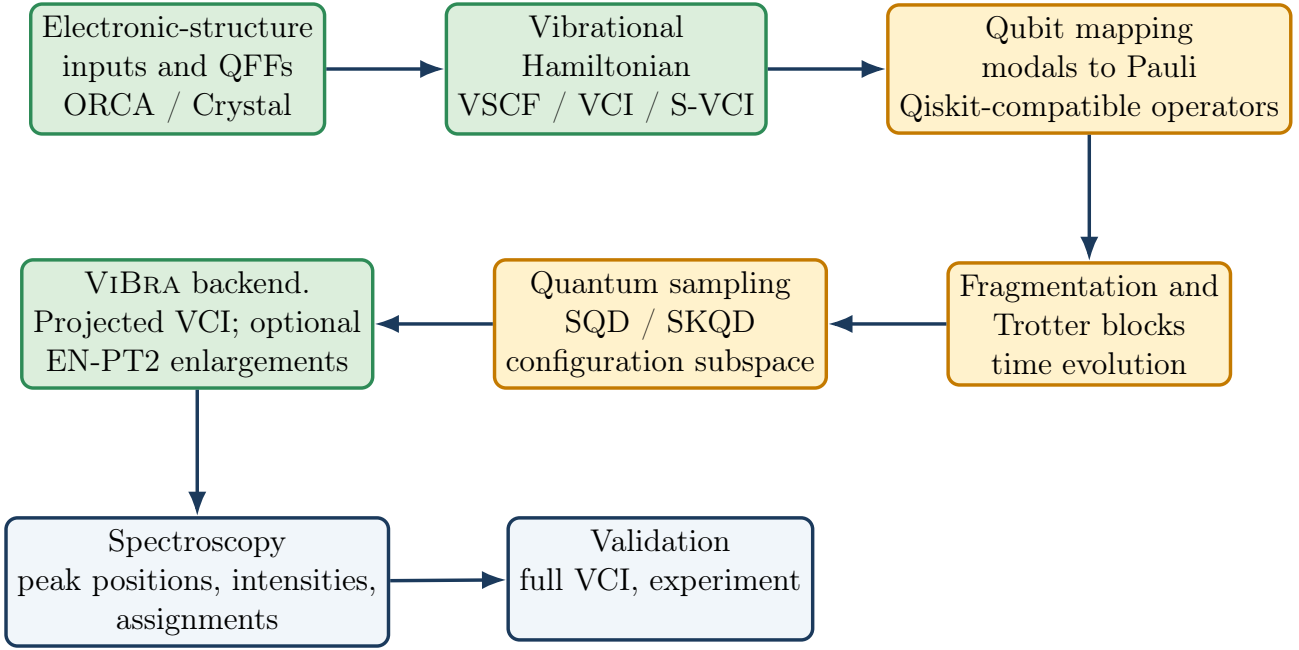

Figure~\ref{fig:quantum_vibra_workflow} summarises the proposed workflow. The
starting point is the same as in a conventional anharmonic vibrational
calculation: electronic-structure calculations provide normal-mode
frequencies, cubic and quartic force constants, and dipole derivatives. From
these data, \vibra{} constructs the VSCF modal basis and the corresponding
VCI Hamiltonian. 

For the quantum-facing part of the workflow, the same
vibrational Hamiltonian is mapped to qubit operators using a boson-to-qubit
or modal-to-qubit representation. 
In the compact configuration-index encoding employed here, the $D(M,N_q)=\binom{N_q+M}{M}$ configurations satisfying $\sum_{i=1}^{M}n_i\leq N_q$ are mapped bijectively to computational-basis states of an $n_{\mathrm{qb}}=\lceil\log_2 D(M,N_q)\rceil$-qubit register, with any remaining basis states treated as unused padding.
After mapping, the Hamiltonian takes the Pauli form
\begin{equation}
\Hvib
\longrightarrow
\hat{H}_{q}
=
\sum_{k} c_k P_k,
\label{eq:pauli_ham}
\end{equation}
where $P_k$ are tensor products of single-qubit Pauli operators. The mapped
Hamiltonian can then be partitioned into mutually commuting blocks of Pauli operators,
\begin{equation}
\hat{H}_{q} = \sum_{\nu} \hat{H}_{\nu},
\label{eq:fragmented_ham}
\end{equation}
so that a real-time propagator can be approximated using a product formula, such as the first-order Suzuki-Trotter decomposition
\begin{equation}
\hat{U}_{\mathrm{Trot}}(t)
=
\left[
\prod_{\nu}
\exp\left(-i\hat{H}_{\nu}t/r\right)
\right]^r
+
\mathcal{O}(t^2/r),
\label{eq:trotter_propagator}
\end{equation}
with $r$ Trotter steps. For the initial state of the evolution, we have used a normalized superposition of the vacuum and all first
single-mode excitations. The resulting circuits can be used either to generate
samples from states relevant to the low-energy vibrational spectrum, or to
estimate a time-domain correlation function directly.

Given a set of measured bit strings $\{\mathbf{b}\}$, the classical post-processing stage decodes each valid string into a vibrational configuration $\bn=\mathcal{D}(\mathbf{b})$. In a binary modal encoding, invalid strings that do not contain one occupied modal per mode can be removed by post-selection or, when appropriate, processed using a configuration-recovery procedure analogous to that introduced in the original SQD work \cite{RobledoMoreno2025SQD}. The quantum-sampled subspace is then

\begin{equation}
\mathcal{S}_{\mathrm{Q}}
=
\left\{
\bn \; \middle| \;
\bn=\mathcal{D}(\mathbf{b}),\;
\mathbf{b}\in\mathrm{samples},\;
c(\mathbf{b})\ge c_{\min}
\right\},
\label{eq:q_sampled_space}
\end{equation}
where $c(\mathbf{b})$ is the observed count of bit string $\mathbf{b}$ and $c_{\min}$ is an optional count threshold. In practice, the sampled space can be formed as the union of configurations obtained from multiple state preparations, time steps, or Krylov-like states. Once $\mathcal{S}_{\mathrm{Q}}$ is defined, no additional quantum measurements are required to evaluate the projected Hamiltonian matrix elements: they are computed classically by \vibra{} from the force field and modal integrals.

The quantum-facing components used in this proof of concept are implemented outside the main \vibra{} code base. In particular, the routines for modal-to-qubit mapping, Pauli-operator construction, and Trotterized time evolution follow the implementation provided in~\cite{QuantumVibMappingRepo,QuantumVibFragmentationRepo}. These external routines are used only to generate the qubit Hamiltonian, fragmented evolution operators, quantum circuits, and simulated samples. The subsequent construction of the projected vibrational Hamiltonian, diagonalization of the sampled configuration space, and evaluation of transition intensities are performed by \vibra{}.

\subsection{Workflow Validation for \ce{H2O}}
\label{validation_qpu_h20}
As a proof-of-concept demonstration, we applied this workflow to the water molecule. Water is a convenient initial test system: its three vibrational modes keep the full, truncated VCI problem exactly tractable, while still incorporating nontrivial anharmonic couplings and infrared-active fundamentals. We performed the calculation for several values of truncation of the configuration space $N_q$, varying it from 2 to 12, with the full VCI(12) calculation serving as our reference. All \vibra{} calculations in this section were performed using: \texttt{NEXPAN}=15, \texttt{CVGSCF}=6. We used the same semiquartic force-field data (wB97X-D4/aug-cc-pVTZ) as in the validation section (Steps 3 and 4).

All quantum-facing calculations in this study were performed using a
simulator; no quantum hardware was used. For each value of $N_q$, we prepared a different quantum circuit and we evolved them with 4 Trotter steps separated by a dimensionless time $\delta=0.1$, or a physical time step of $0.531\,ps$, and each circuit was sampled 1000 times. The number of qubits, as well as other relevant information on the circuits is provided in \autoref{tab:gatecounts}.

\begin{table}[htbp]
\centering
\small
\caption{Gate counts for the bosonic SKQD circuits}
\label{tab:gatecounts}
\begin{tabular}{lrrr}
\hline
$N_q$ & Qubits & 1 qubit gates & 2 qubit gates \\
\hline
2  & 4 & 1421    & 494    \\
3  & 5 & 5887    & 2142   \\
4  & 6 & 23902   & 8894   \\
5  & 6 & 23895   & 8894   \\
6  & 7 & 96274   & 36222  \\
7  & 7 & 96283   & 36222  \\
8  & 8 & 386446  & 146174 \\
9  & 8 & 386414  & 146174 \\
10 & 9 & 1548396 & 587262 \\
11 & 9 & 1548438 & 587262 \\
12 & 9 & 1548446 & 587262 \\
\hline
\end{tabular}
\end{table}


The sampled bit strings were decoded into vibrational configurations, and the union of the configurations
over all time steps defines the quantum-generated seed space
$\mathcal{S}_{\mathrm{Q}}(N_q)$. The results reported below correspond to
\path|\plot_water_qpu_bench\bosonic_selected_states_water_|\(N_q\)\path|_quanta|
in the database.\cite{QuantumVibData}

The sampled configurations were supplied to \vibra{} in list-reference mode.
With \texttt{MAXSCI 0 list}, \vibra{} constructs and diagonalizes the
Hamiltonian directly in $\mathcal{S}_{\mathrm{Q}}$. With
\texttt{MAXSCI} $m$ \texttt{list}, where $m=1$, 2, or 3, the sampled list is
instead used as the reference for an optional classical enlargement. External configurations are ranked for each targeted root using the Epstein--Nesbet second-order perturbative importance measure described in Section~\ref{sec:SCI}. The highest-ranked configurations are added to the quantum seed, duplicates are removed, and the Hamiltonian is variationally rediagonalized in the resulting space,
\[
\mathcal{S}^{(m)}
=
\mathcal{S}_{\mathrm{Q}}
\cup
\mathcal{S}_{\mathrm{EN}}^{(m)}.
\]
Table~\ref{tab:tab_water_sattes} reports the total number of states obtained in each case. 

Thus, the quantum-facing routine generates the initial seed configurations,
while \vibra{} can either analyze that seed directly or enlarge it through
classical EN-PT2-guided selection. The EN-PT2 quantities are used only to rank
additional configurations; they are not added as perturbative corrections to
the reported energies, which are eigenvalues obtained by variational
rediagonalization of the final projected Hamiltonian.


For
$N_q=2$--4, the sampling procedure retained every
configuration in the corresponding VCI space. Those calculations therefore
do not test configuration filtering and are omitted from the following
discussion. At $N_q=5$, 51 of the 56 available configurations
were retained, and we consequently discuss the results from
$N_q=5$ onward.



\begin{table}[t]
\centering
\small
\setlength{\tabcolsep}{3.8pt}
\caption{Dimensions of the quantum-sampled seed spaces and of the spaces obtained after optional EN-PT2-guided enlargement by \vibra{}. The row with \texttt{MAXSCI}=0 corresponds to the direct use of the externally sampled seed. The remaining rows are quantum-seeded spaces that have been enlarged by \vibra{} and variationally rediagonalized. The VCI row gives the maximum number of states for a given $N_q$, which is $\binom{N_q+3}{3}$, where 3 is the number of modes for water. Note that the enlarged spaces exceed their respective counts because the enlargement was performed, for all $N_q$ (in the qubit mapping), using \texttt{NQUANT}=12 in \vibra{}.}
\label{tab:tab_water_sattes}
\begin{tabular}{lrrrrrrrr}
\toprule
$N_q$  & 5 & 6 & 7 & 8 & 9 & 10 & 11 & 12 \\
\midrule
VCI
  & 56 & 84 & 120 & 165 & 220 & 286 & 364 & 455 \\
Quantum seed, \texttt{MAXSCI}=0
  & 51 & 65 & 82 & 74 & 74 & 61 & 62 & 53 \\
ViBra-enlarged, \texttt{MAXSCI}=1
  & 84 & 105 & 125 & 118 & 118 & 100 & 101 & 86 \\
ViBra-enlarged, \texttt{MAXSCI}=2
  & 107 & 125 & 149 & 136 & 136 & 120 & 122 & 106 \\
ViBra-enlarged, \texttt{MAXSCI}=3
  & 115 & 142 & 163 & 155 & 155 & 136 & 140 & 118 \\
\bottomrule
\end{tabular}
\end{table}


Table~\ref{tab:tab_water_sattes} shows that the absolute size of the sampled
seed is nonmonotonic with $N_q$: the directly sampled lists
contain between 51 and 82 configurations. Relative to the available space,
however, the retained fraction decreases from 51/56 (91.1\%) at
$N_q=5$ to 53/455 (11.6\%) at
$N_q=12$. Increasing the sampling cutoff therefore does not,
under the present sampling protocol, guarantee a larger seed or a more
complete representation of a target eigenstate. The identity and connectivity
of the sampled configurations are at least as important as their number.

\begin{figure}[hbt!]
    \centering
    \setcounter{subfigure}{0}

    \begin{subfigure}[t]{0.285\textwidth}
        \centering
        \phantomsubcaption\label{fig:energy-a}%
        \includegraphics[width=\linewidth]{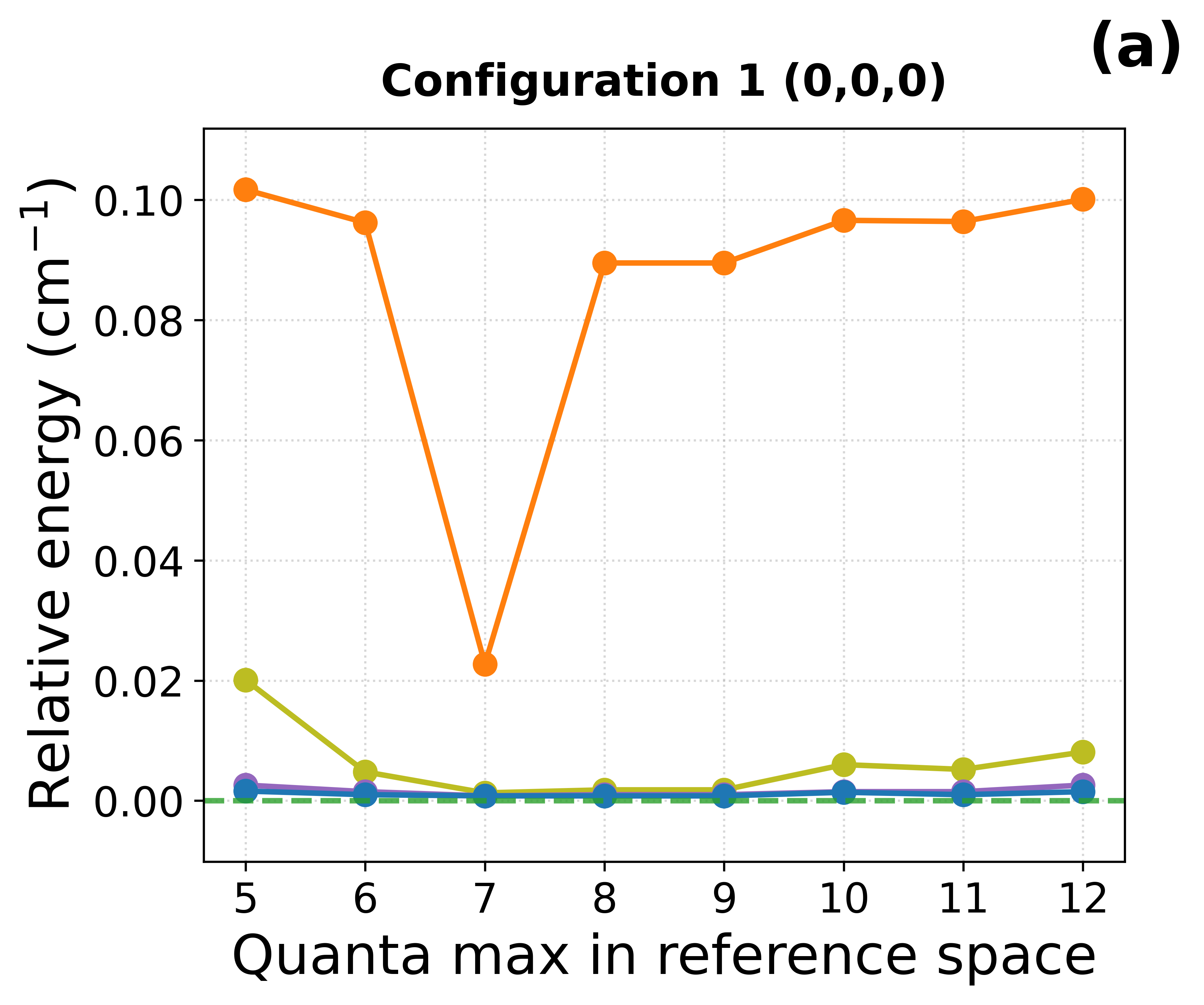}
    \end{subfigure}\hfill
    \begin{subfigure}[t]{0.285\textwidth}
        \centering
        \phantomsubcaption\label{fig:energy-b}%
        \includegraphics[width=\linewidth]{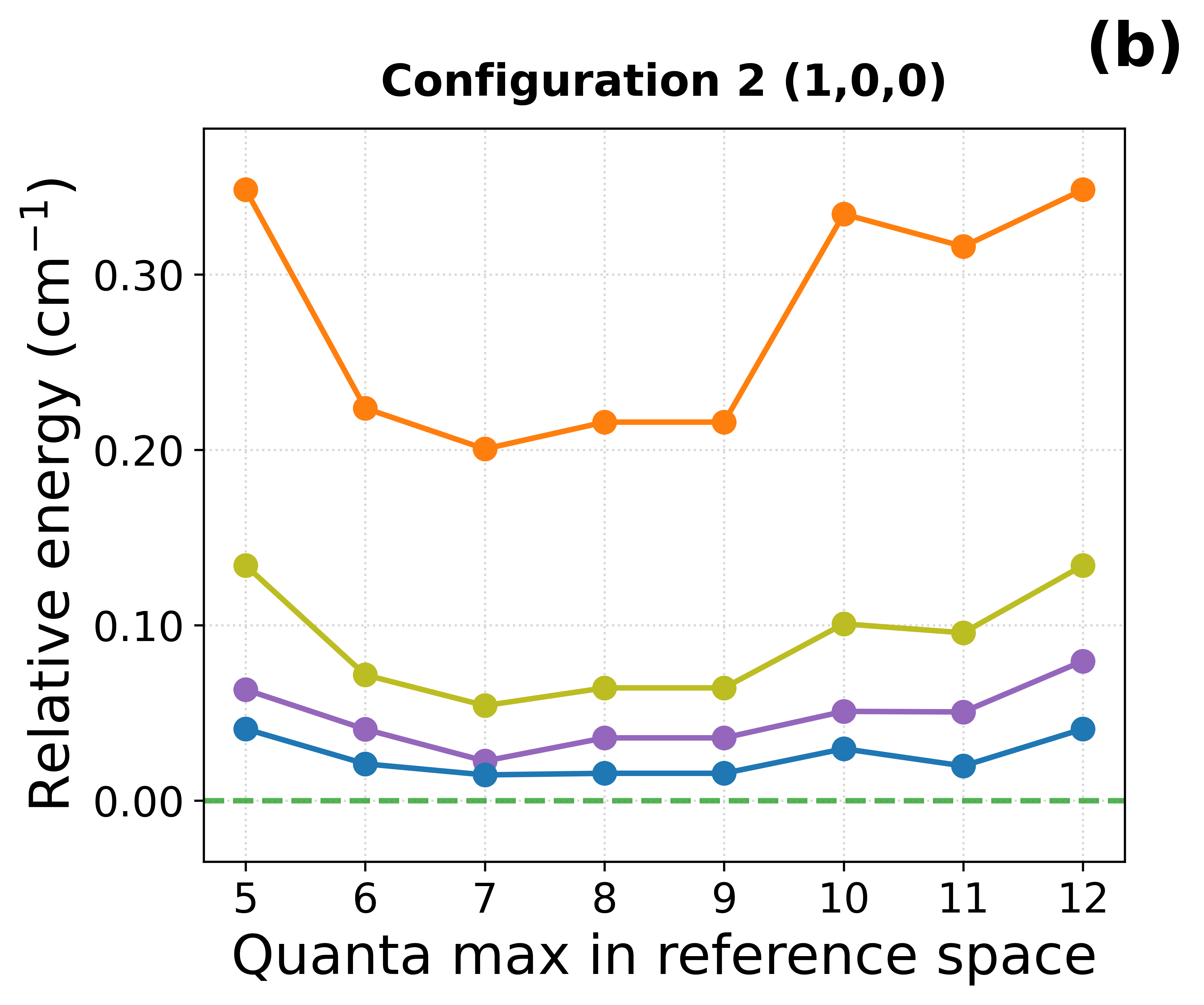}
    \end{subfigure}\hfill
    \begin{subfigure}[t]{0.285\textwidth}
        \centering
        \phantomsubcaption\label{fig:energy-c}%
        \includegraphics[width=\linewidth]{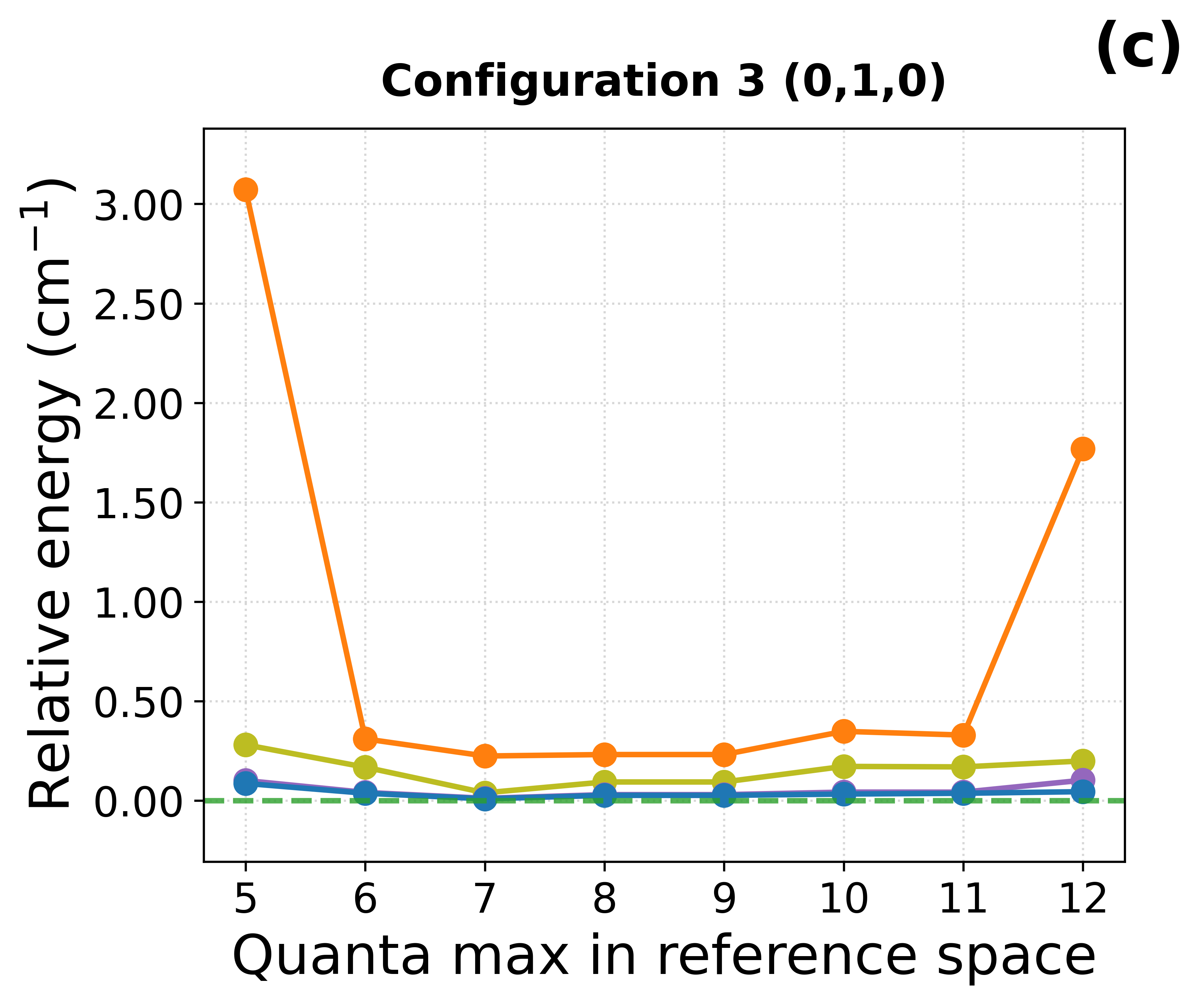}
    \end{subfigure}

    \begin{subfigure}[t]{0.285\textwidth}
        \centering
        \phantomsubcaption\label{fig:energy-d}%
        \includegraphics[width=\linewidth]{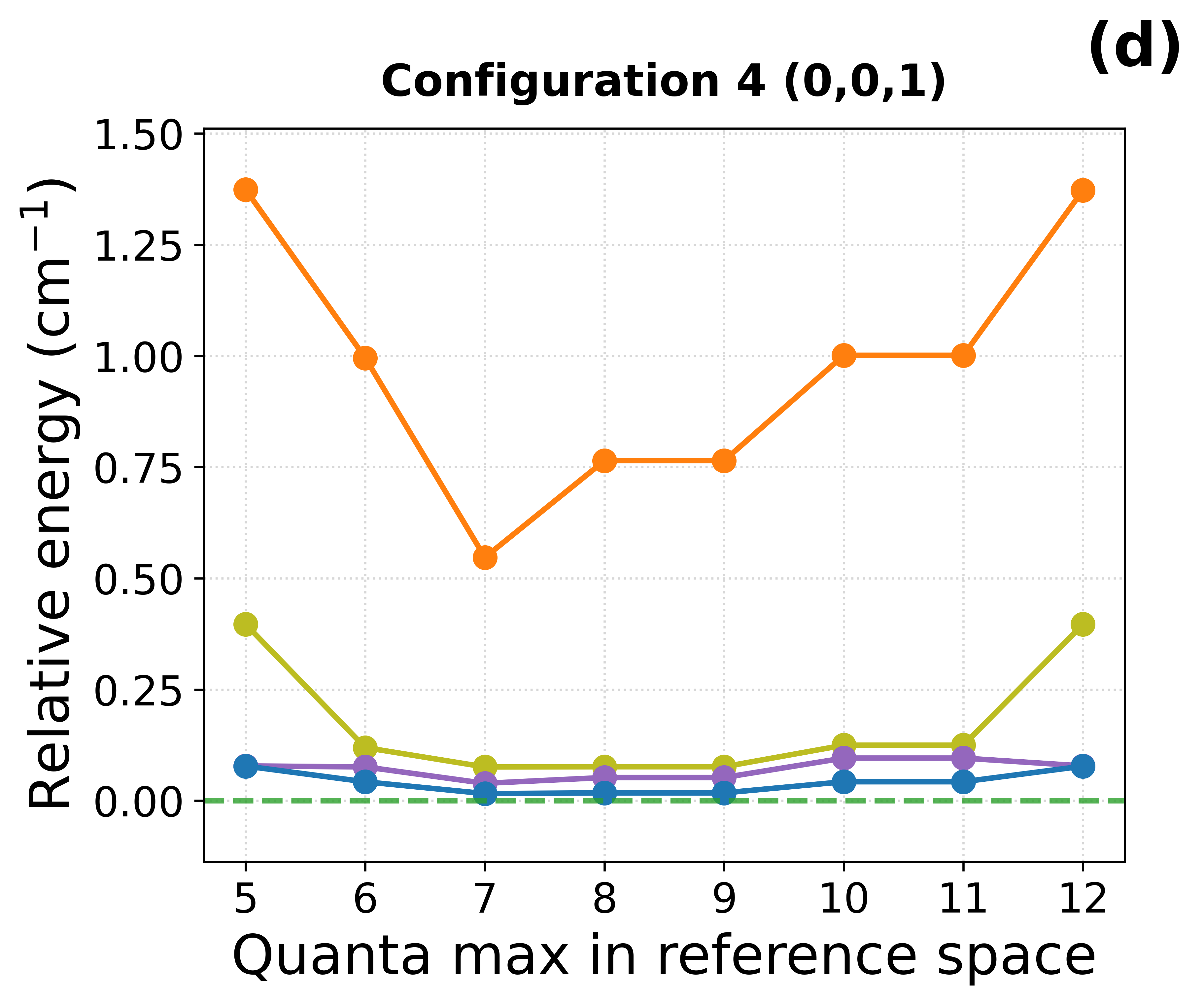}
    \end{subfigure}\hfill
    \begin{subfigure}[t]{0.285\textwidth}
        \centering
        \phantomsubcaption\label{fig:energy-e}%
        \includegraphics[width=\linewidth]{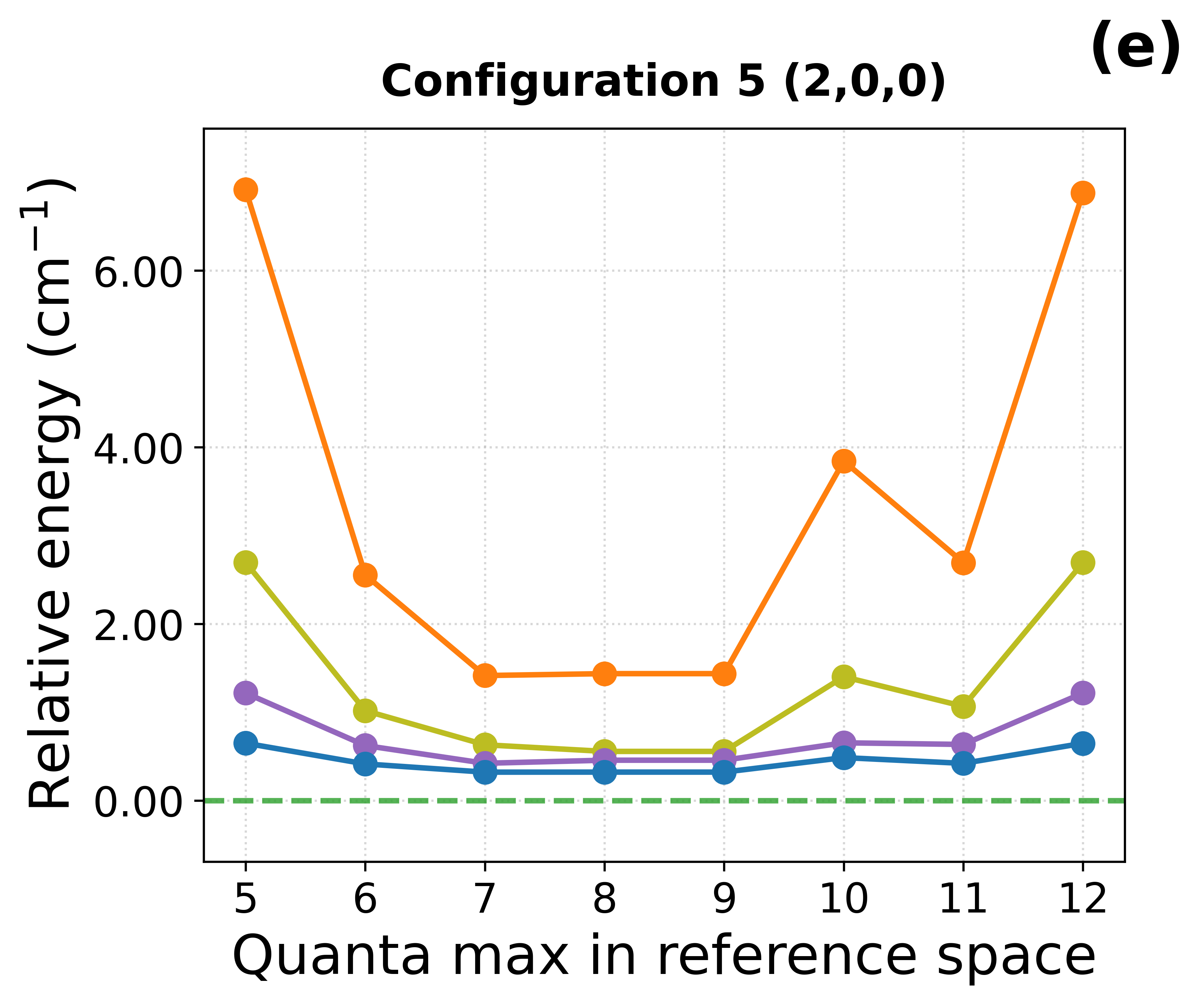}
    \end{subfigure}\hfill
    \begin{subfigure}[t]{0.285\textwidth}
        \centering
        \phantomsubcaption\label{fig:energy-f}%
        \includegraphics[width=\linewidth]{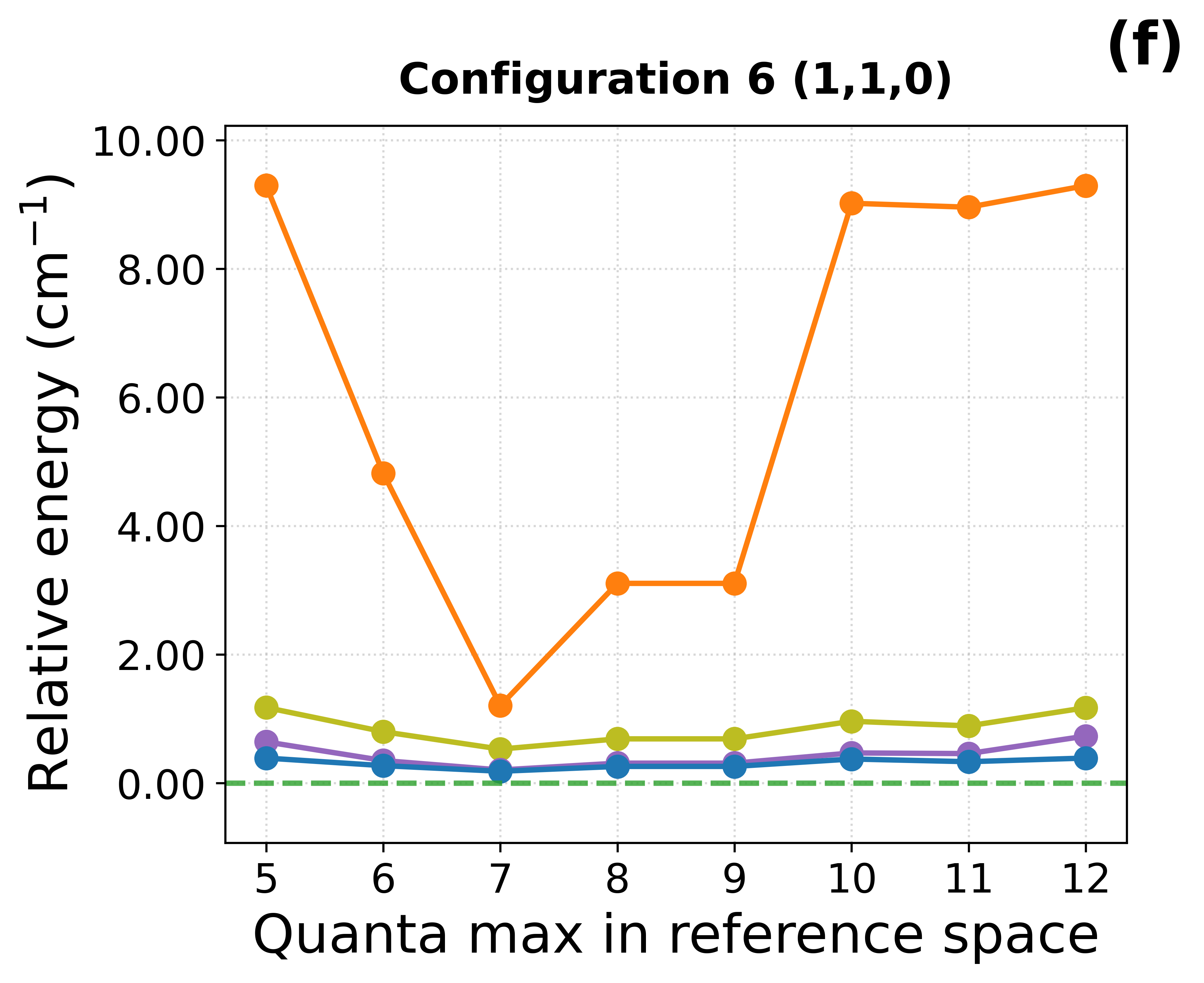}
    \end{subfigure}

    \begin{subfigure}[t]{0.285\textwidth}
        \centering
        \phantomsubcaption\label{fig:energy-g}%
        \includegraphics[width=\linewidth]{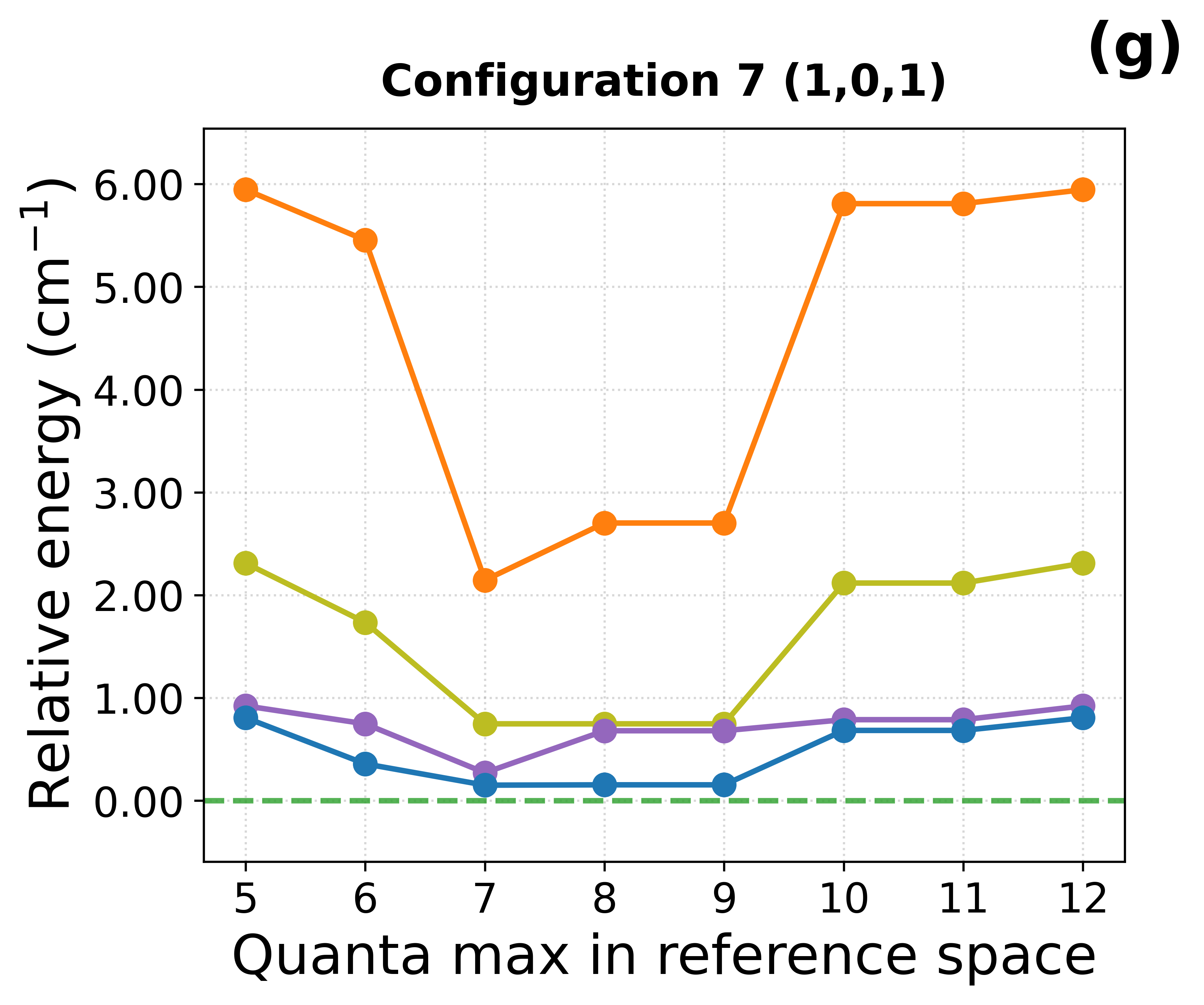}
    \end{subfigure}\hfill
    \begin{subfigure}[t]{0.285\textwidth}
        \centering
        \phantomsubcaption\label{fig:energy-h}%
        \includegraphics[width=\linewidth]{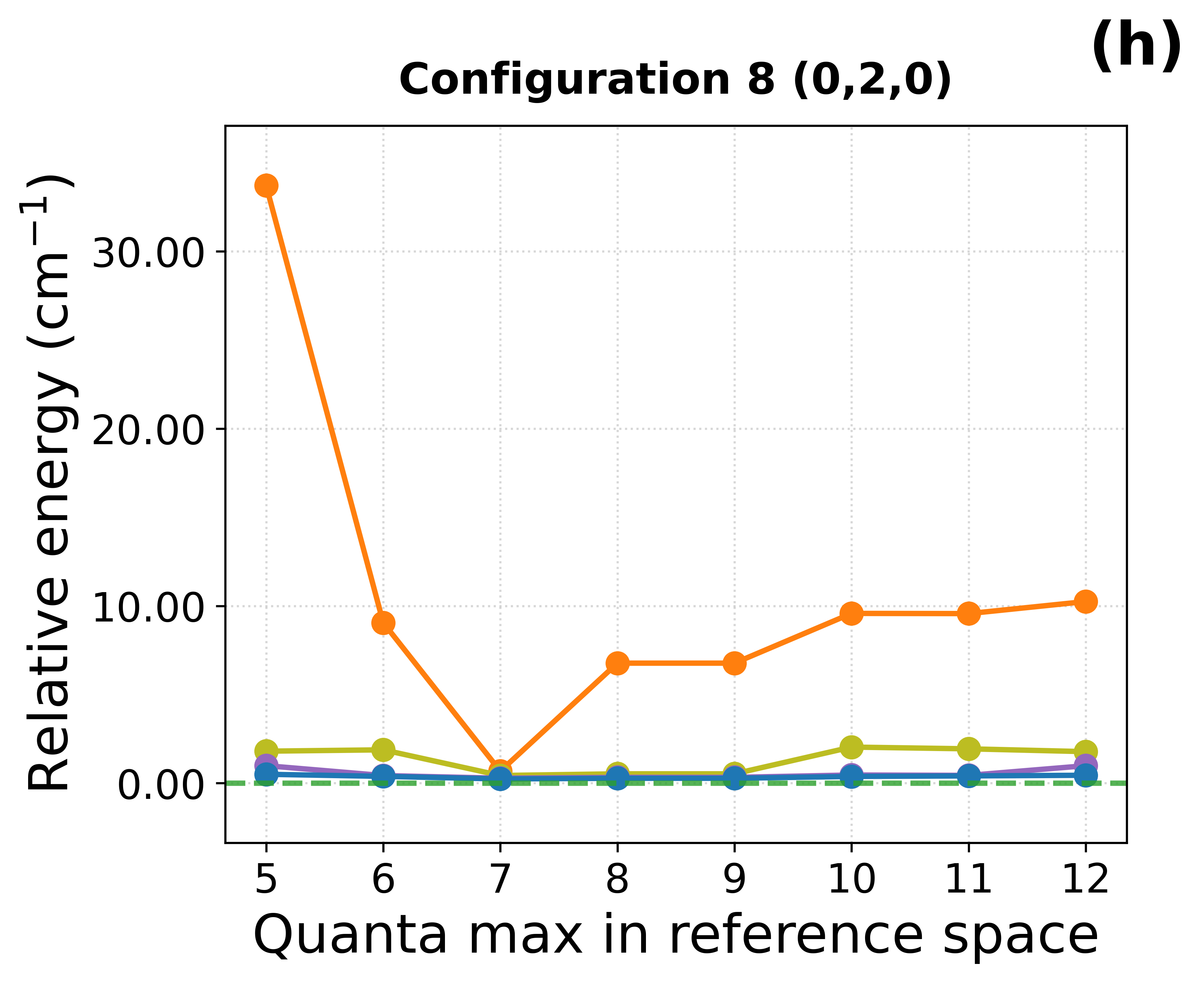}
    \end{subfigure}\hfill
    \begin{subfigure}[t]{0.285\textwidth}
        \centering
        \phantomsubcaption\label{fig:energy-i}%
        \includegraphics[width=\linewidth]{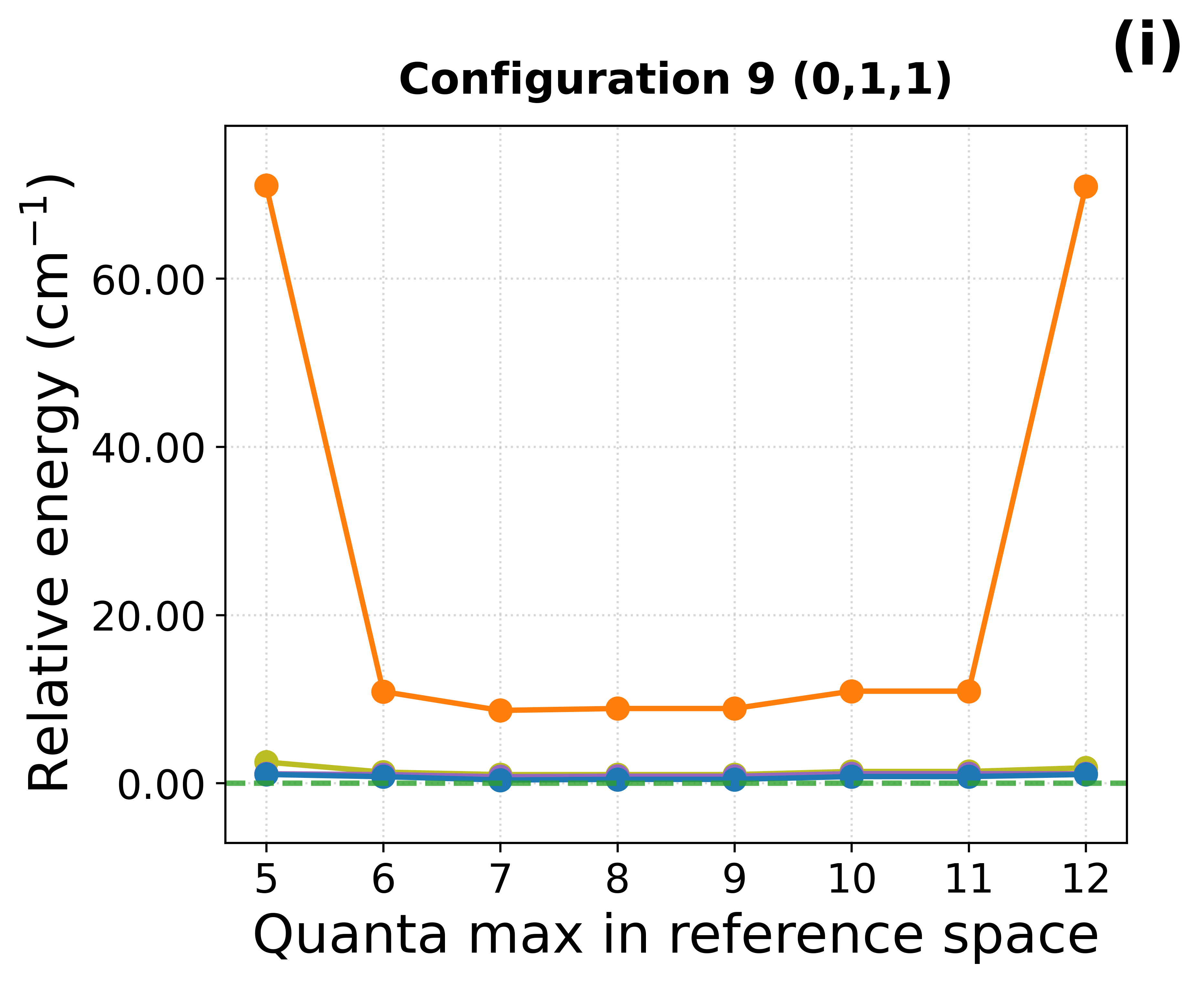}
    \end{subfigure}

    \makebox[\textwidth][c]{%
        \begin{subfigure}[c]{0.285\textwidth}
            \centering
            \phantomsubcaption\label{fig:energy-j}%
            \includegraphics[width=\linewidth]{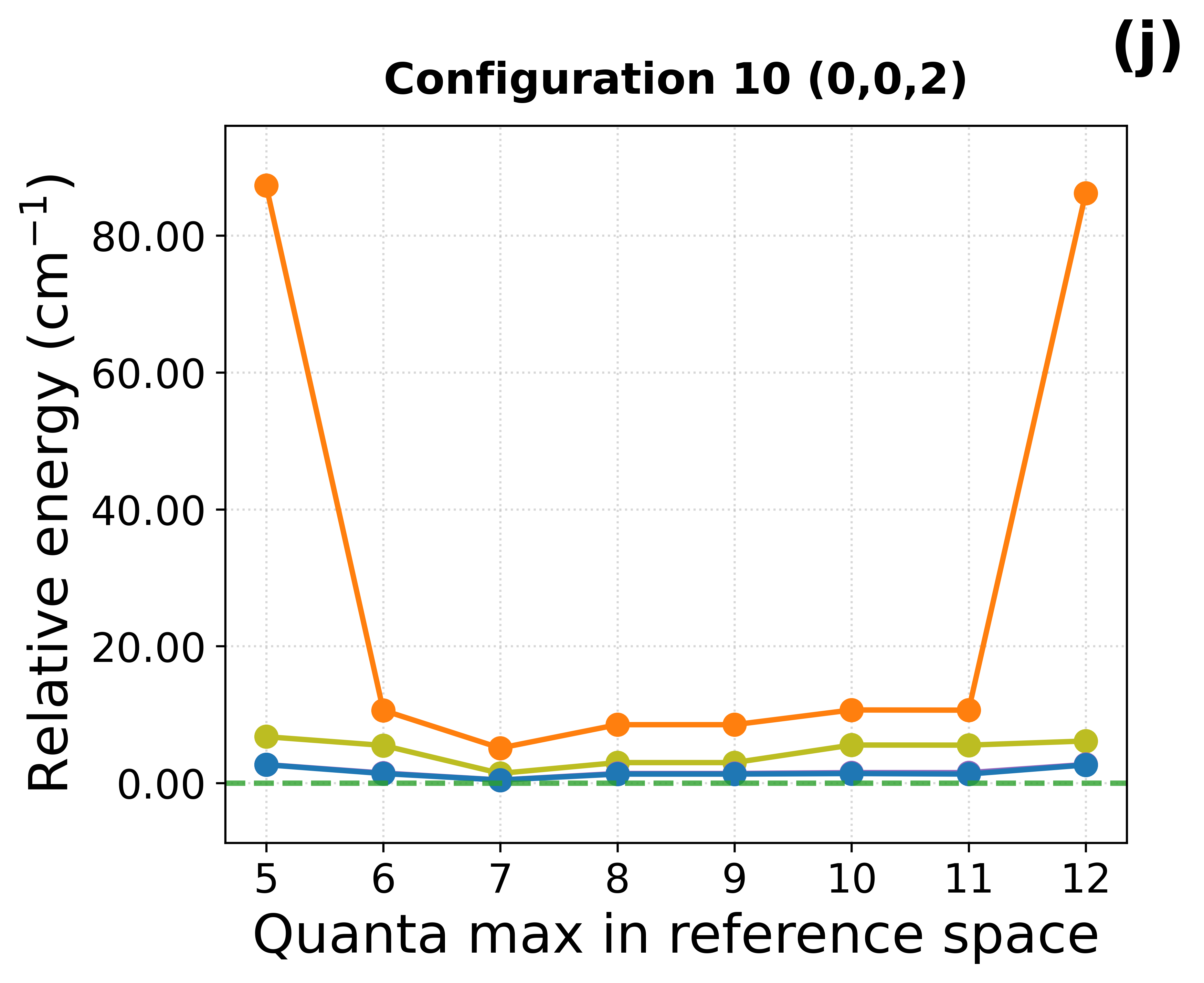}
        \end{subfigure}%
        \hspace{0.05\textwidth}%
        \begin{minipage}[c]{0.13\textwidth}
            \centering
            \includegraphics[width=\linewidth]{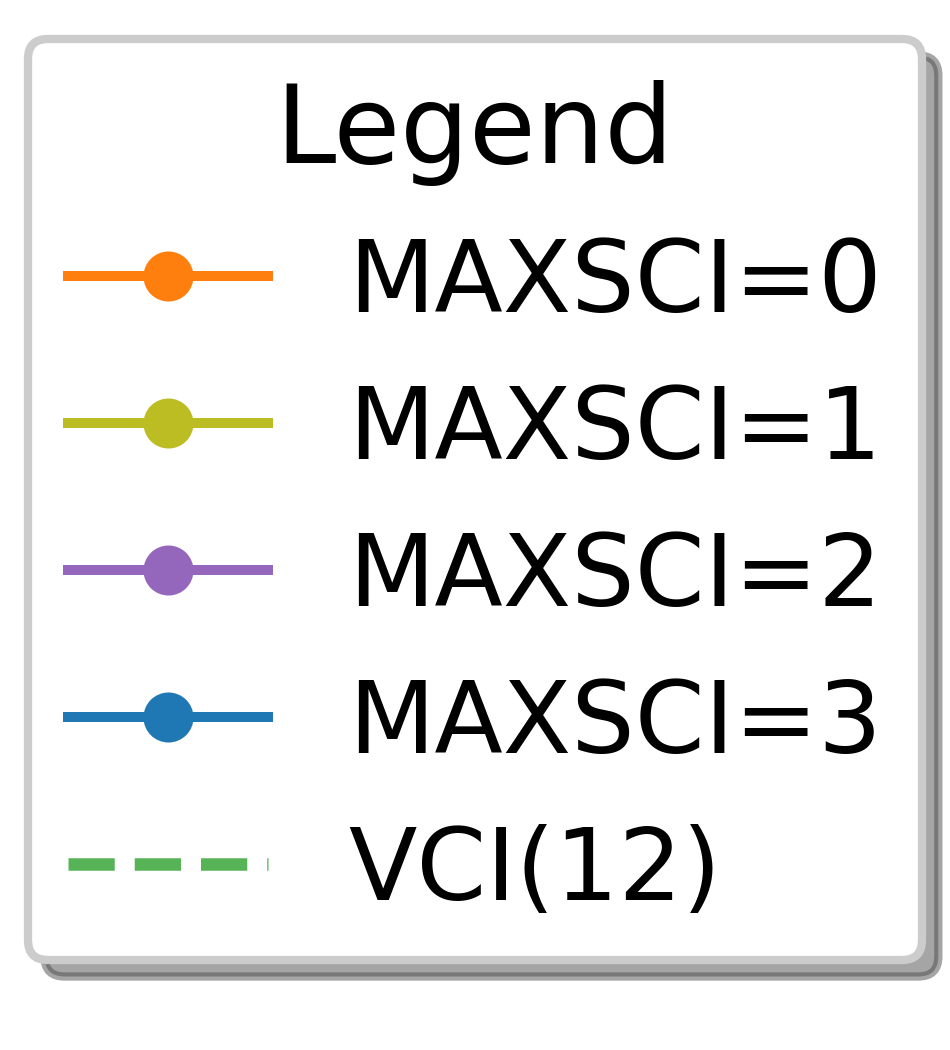}
        \end{minipage}%
    }

    \caption{Signed energy deviation, for the water
molecule as a function of the maximum total quanta
$N_q=5$--12 used to generate the quantum-sampled seed. Panel (a) shows the ground state; panels (b)--(j) show the nine states dominated by the indicated configurations. \texttt{MAXSCI}=0 denotes direct variational diagonalization of the sampled seed, while \texttt{MAXSCI}=1--3 denote EN-PT2-guided enlargement of that seed by \vibra{} followed by variational rediagonalization. EN-PT2 is used only for configuration ranking; no perturbative correction is added to the reported energies.}
    \label{fig_energy}
\end{figure}

To assess the physical quality of the resulting spaces, we considered the
ground state together with the nine states whose dominant reference
configurations span the fundamentals, overtones, and combination bands through
two quanta. Figure~\ref{fig_energy} reports the signed total-energy deviation. For each indicated reference configuration, we identified the projected-space
root to which that configuration contributes the largest wavefunction
coefficient.

Direct diagonalization of the quantum seed reproduces the zero-point energy to
approximately 0.1~cm$^{-1}$ across the cutoff range and generally performs
well for the low-energy one-quantum states. Larger and more variable
deviations occur for the two-quantum manifold. In particular, the direct-space
deviations reach approximately 70--80~cm$^{-1}$ for the configurations shown
in Figs.~\ref{fig_energy}\subref{fig:energy-i} and
\ref{fig_energy}\subref{fig:energy-j}. These large deviations persist at both
the lower and upper ends of the sampling-cutoff range.

The agreement with the VCI benchmarks improves markedly after EN-PT2-guided configuration-space enlargement by \vibra{}, at the cost of larger projected Hamiltonians and correspondingly longer calculations. Even in the worst case, the \texttt{MAXSCI}=3 list (163 of 455 states) comprises only 36\% of the full VCI(12) space, a substantial reduction. Already at \texttt{MAXSCI}=1, errors that previously exceeded 80 cm$^{-1}$ drop below 5 cm$^{-1}$ for all ten states considered. Moving from \texttt{MAXSCI}=2 to 3 tightens this further, to roughly 1 cm$^{-1}$, showing an excellent agreement with the reference. We note that Figure~\ref{fig_energy}h is the only state for which the $N_q=5$ and $12$ results differ significantly once $N_q$ reaches 12, with the energy discrepancy dropping from roughly 30 to 10 cm$^{-1}$. This might suggest that the unenlarged space is of higher quality at larger quanta limits; however, this is not observed as a general trend, and no broader conclusion should be drawn from this single case. 

\begin{figure}[p]
    \centering
    \setcounter{subfigure}{0}

    \begin{subfigure}[t]{0.485\textwidth}
        \centering
        \phantomsubcaption\label{fig:composition-a}%
        \includegraphics[width=\linewidth]{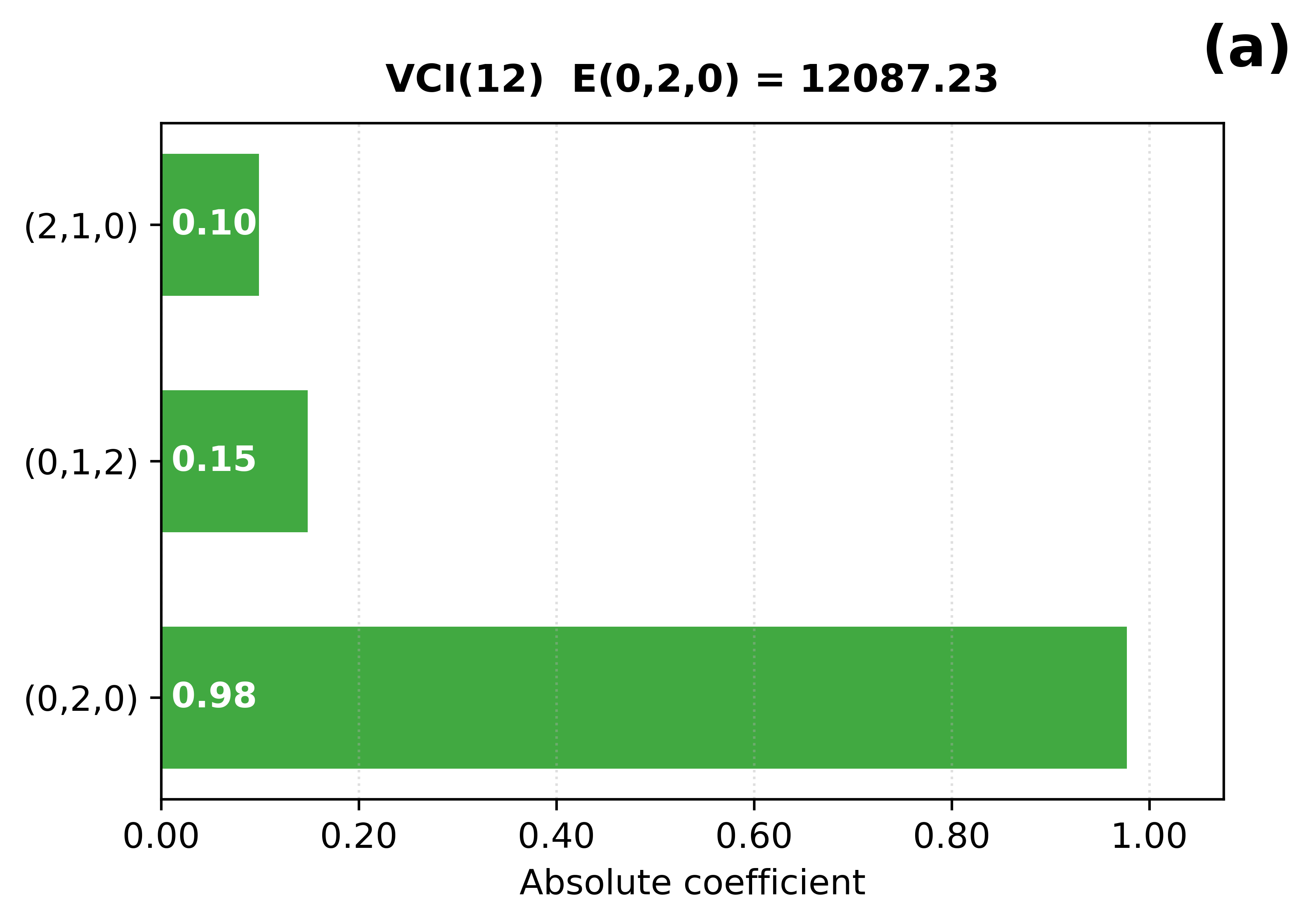}
    \end{subfigure}\hfill%
    \begin{subfigure}[t]{0.485\textwidth}
        \centering
        \phantomsubcaption\label{fig:composition-b}%
        \includegraphics[width=\linewidth]{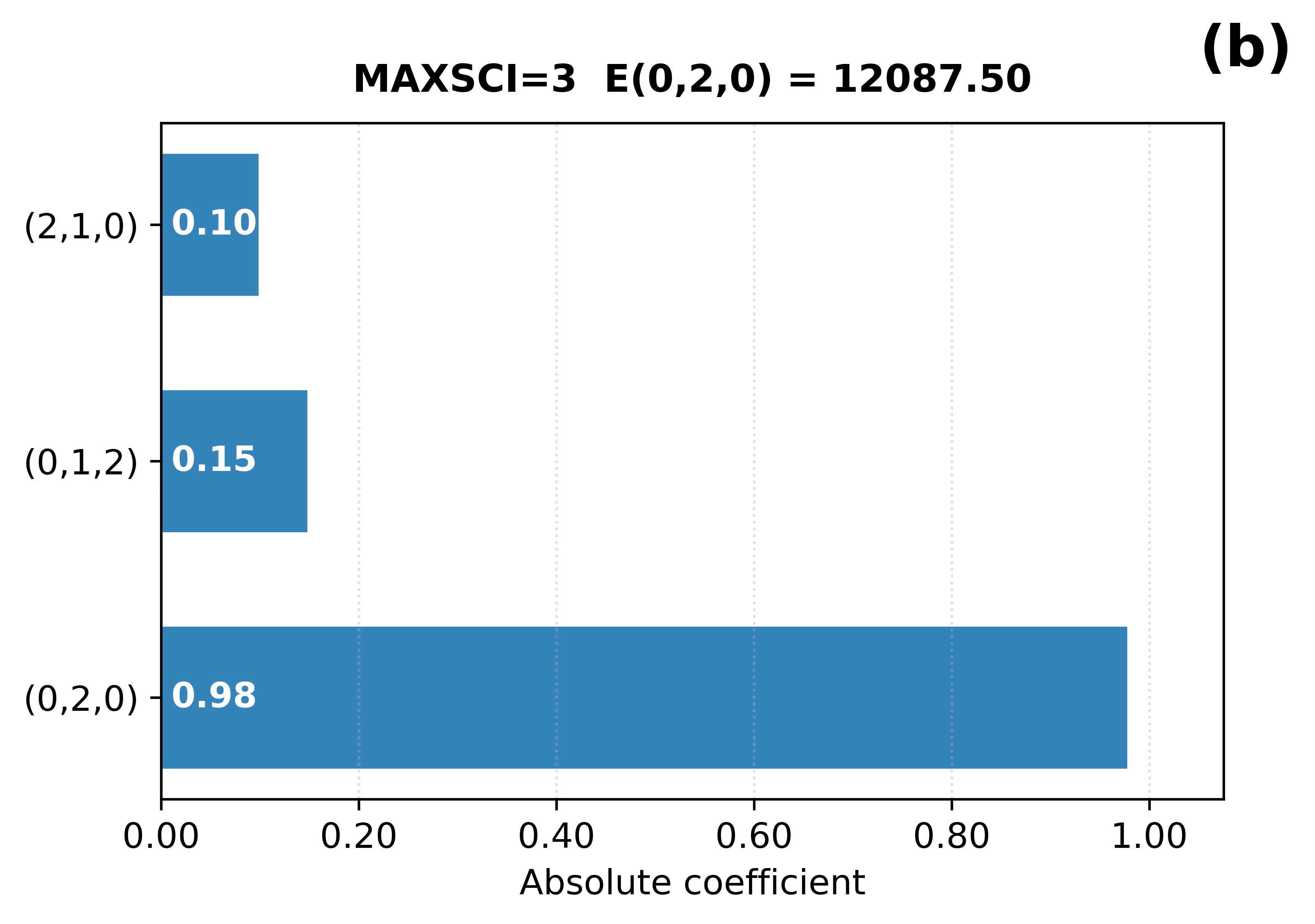}
    \end{subfigure}

    \par\medskip

    \begin{subfigure}[t]{0.485\textwidth}
        \centering
        \phantomsubcaption\label{fig:composition-c}%
        \includegraphics[width=\linewidth]{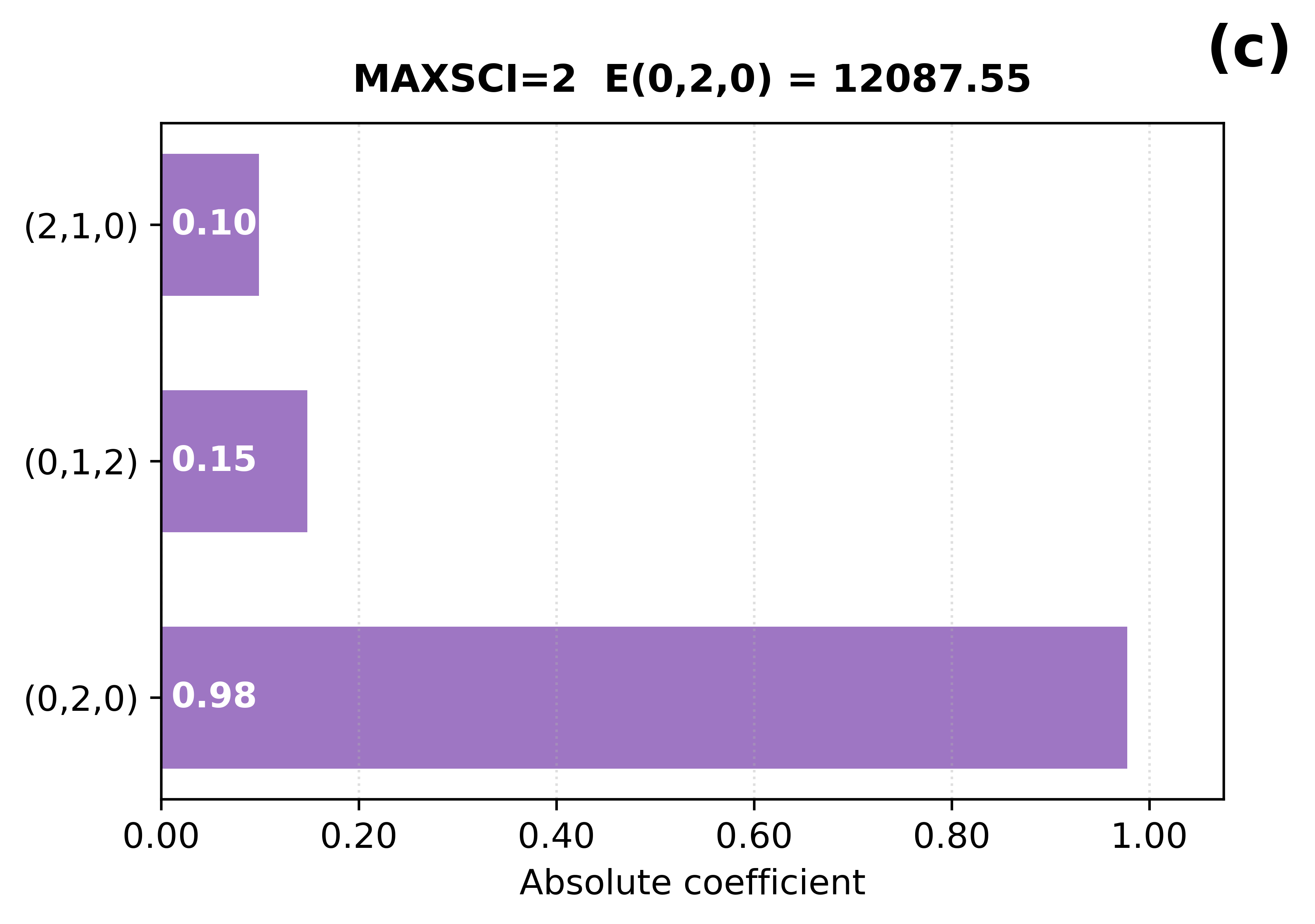}
    \end{subfigure}\hfill%
    \begin{subfigure}[t]{0.485\textwidth}
        \centering
        \phantomsubcaption\label{fig:composition-d}%
        \includegraphics[width=\linewidth]{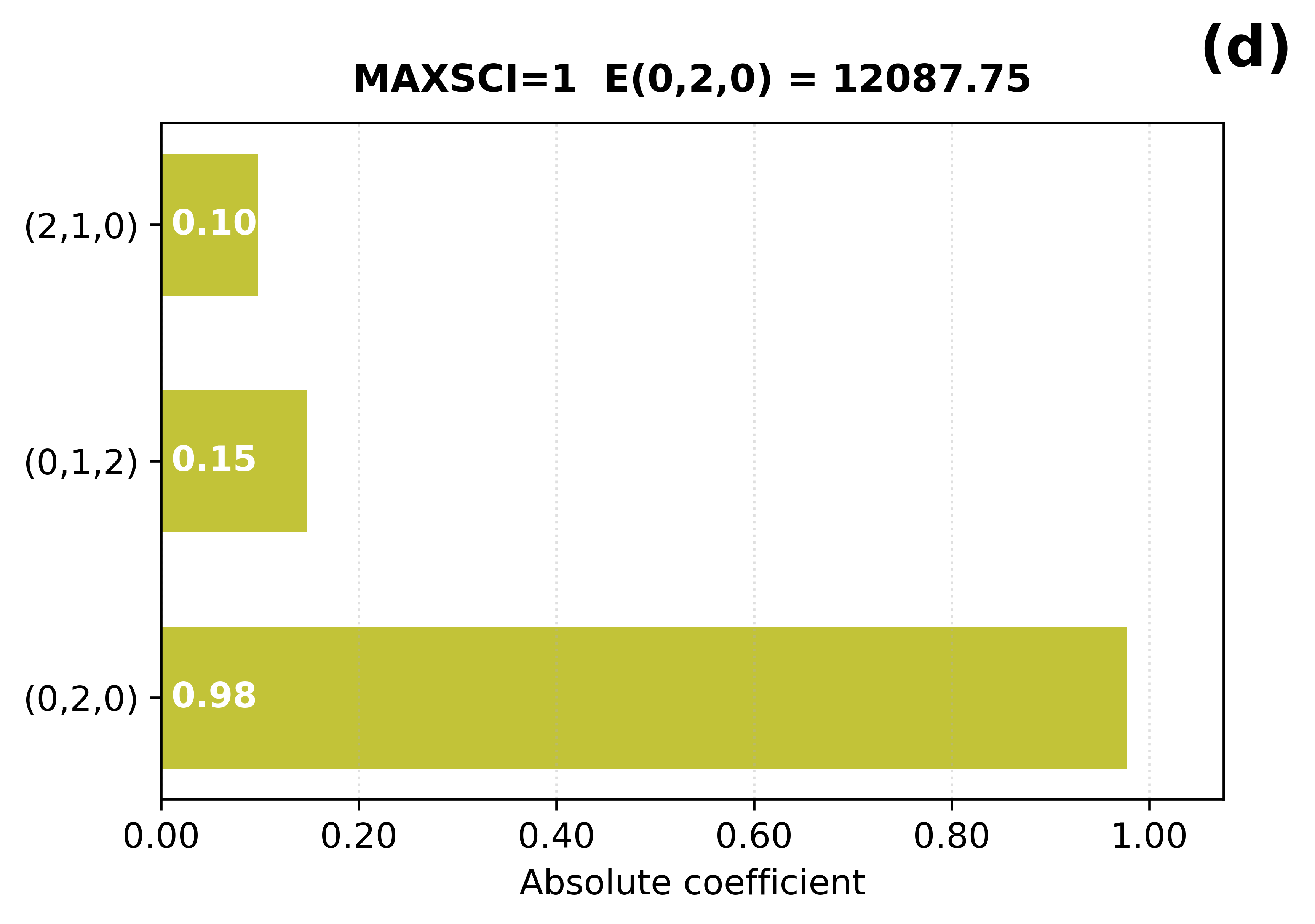}
    \end{subfigure}

    \par\medskip

    \begin{subfigure}[t]{0.485\textwidth}
        \centering
        \phantomsubcaption\label{fig:composition-e}%
        \includegraphics[width=\linewidth]{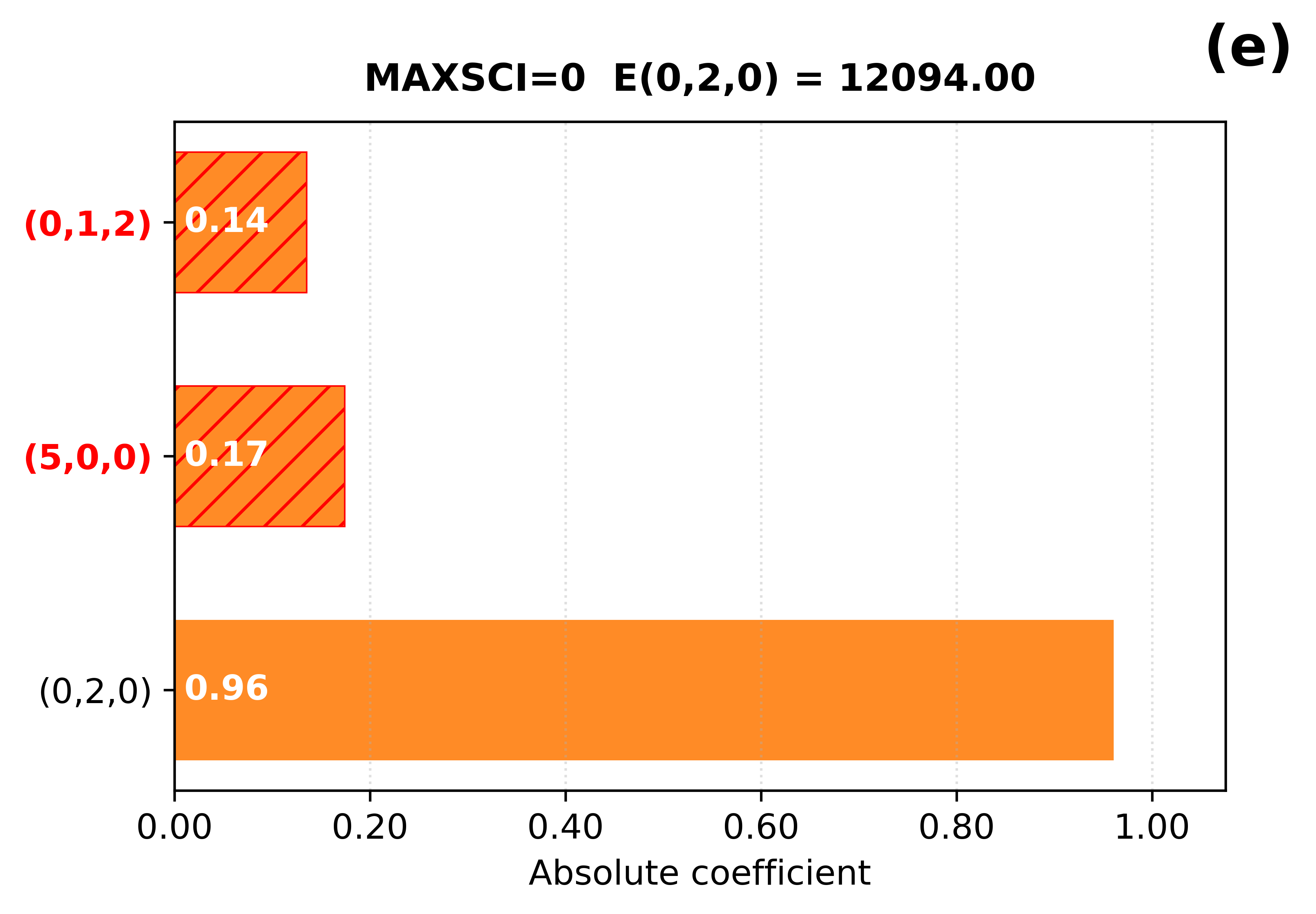}
    \end{subfigure}
    \caption{Absolute dominant configuration coefficients for the $(0,2,0)$ state at sampling cutoff $N_q=8$. Panel (a) shows the full VCI(12) reference; panels (b-d)  show the ViBra-enlarged spaces with \texttt{MAXSCI}=1 to 3, respectively. (e) shows  direct diagonalization of the quantum-sampled seed with \texttt{MAXSCI}=0, highlighting in red the wrong coupling of states. The recovery of the reference composition at \texttt{MAXSCI}=1--3 illustrates the effect of classical EN-PT2-guided configuration-space enlargement after quantum sampling.}
    \label{fig_comp}
\end{figure}

An important question is whether the improved energetic agreement alone is sufficient to validate the SKQD+PT2 approach. Figure~\ref{fig_comp} is instructive here. The wavefunction composition analysis in Fig.~\ref{fig_comp} helps to elucidate a classic truncated selected CI problem. In the VCI(12) eigenvector, $(0,2,0)$ is dominant and $(0,1,2)$ is its largest secondary contribution. In the directly sampled space, $(5,0,0)$ instead becomes the largest secondary contribution. Even with \texttt{MAXSCI}=1, it was sufficient to recover the composition of the wavefunction. It is important to note that the state $(0,1,2)$ is present in all calculations, spanning \texttt{MAXSCI} from 0 to 3. This does not mean that \vibra{} evaluates the retained off-diagonal Hamiltonian elements incorrectly. Rather, the incomplete configuration space omits contributions needed to reproduce the effective mixing pattern of the full calculation. With \texttt{MAXSCI}=1--3, the dominant coefficients closely recover the VCI(12) composition.




The immediate conclusion is that the unenlarged quantum-sampled seed should be used with caution: while it may yield deceptively favorable energies, it can also introduce spurious state mixing or apparent resonances. We would also expect an incomplete space to under-represent peak splitting where it should occur. We therefore emphasize the value of inspecting wavefunction composition, peak intensities, and splittings alongside the energies.

These results establish the intended role of \vibra{} in a hybrid quantum--classical vibrational workflow. The simulated quantum-facing routine supplies an initial set of vibrational configurations, $\mathcal{S}_{\mathrm{Q}}$. With \texttt{MAXSCI 0 list}, \vibra{} constructs and diagonalizes the projected Hamiltonian directly in this externally sampled space. With \texttt{MAXSCI}>0 in list-reference mode, the same sampled set is used as the reference for \vibra{}'s EN-PT2-guided configuration selection, producing a quantum-seeded but classically enlarged subspace. In both cases, the reported energies and eigenvectors are obtained through variational
rediagonalization in the final configuration space.

The present \ce{H2O} calculations are intended as workflow and interface validation. They show that quantum-generated seed configurations can be
decoded, supplied to \vibra{}, analyzed directly, or systematically enlarged
through its selected-VCI machinery. All quantum-facing calculations were performed using a simulator, and no execution on quantum hardware was undertaken. Accordingly, these results do not constitute a benchmark of SQD
or SKQD, a study of shot complexity or scaling, or a claim of quantum advantage. Future work should examine configuration-recovery strategies\cite{RobledoMoreno2025SQD,Barison_2025,Barroca2026SurfaceReactionsSQD,Duriez2026BandGapsSQD},
alternative SQD/SKQD sampling protocols, circuit and fragmentation optimization\cite{Malpathak2025TrotterVibrational}, alternative initial states, hardware execution, shot requirements, scaling with molecular size and spectral congestion, and the conditions under which a quantum advantage might eventually become possible.

\newpage

\newpage

\clearpage

\section{Conclusion}
\label{sec:conclusion}


The \vibra{} package delivers a complete, efficient, and user friendly toolchain for anharmonic vibrational spectroscopy at the quartic force field level. By integrating configuration selection via S-VCI with symmetry exploitation via SA-VCI, it extends the reach of accurate VCI calculations to larger molecular systems without sacrificing physical rigour. The inclusion of second order dipole derivatives guarantees reliable intensities across the full vibrational spectrum, while the accompanying GUI and viewer make the results immediately interpretable and directly comparable with experiment. The numerical correctness of \vibra{} was established through a four-tier benchmark. The VSCF and full VCI modules were validated against \textsc{Crystal23} for water, achieving agreement within $0.001\,\mathrm{cm}^{-1}$ and $0.01\,\mathrm{cm}^{-1}$, respectively. The SA-VCI and S-VCI modules were validated against \vibra{}'s own full VCI calculations across a test set of five molecular point groups; SA-VCI reproduces full VCI to numerical precision, while S-VCI converges variationally to sub $5\,\mathrm{cm}^{-1}$ accuracy at substantially reduced cost. Taken together, these benchmarks confirm that \textsc{ViBra} is numerically correct and physically faithful.

The selected-VCI infrastructure in \vibra{} also provides a natural interface to quantum-centric algorithms for anharmonic vibrational spectroscopy. In such workflows, a quantum-facing routine supplies samples that define a compact seed space, while \vibra{} either analyzes that space directly or enlarges it through EN-PT2-guided configuration selection before constructing and diagonalizing the projected Hamiltonian and evaluating transition observables. The \ce{H2O} proof of concept validates this two-stage interface by combining qubit mapping, Hamiltonian fragmentation, and simulated
time-evolution sampling with \vibra{} configuration-space refinement and
projected diagonalization. This simulator-based demonstration does not constitute a hardware benchmark or imply quantum advantage, but it serves as motivation for \vibra{} and its capabilities, as well as for the development of future quantum-centric workflows focused on vibrational problems.

In terms of implementation, there are several natural directions in which the current framework can be extended. Although the current focus is on quartic force fields, the code architecture is general and higher order expansions of the potential energy surface can be added straightforwardly. Similarly, the method can in principle be pushed toward still larger systems, though such applications have not yet been tested. On the algorithmic side, the current SA-VCI implementation is restricted to Abelian point groups with one dimensional irreducible representations, and extending it to non-Abelian groups with degenerate irreps is planned for a future release. 

Moreover, a key bottleneck in VCI calculations is the dramatic growth of the Hamiltonian matrix dimension with the number of modes and excitation quanta, which rapidly makes memory the limiting factor. Because the Hamiltonian matrix becomes increasingly sparse as the vibrational space grows, on the fly diagonalizers that exploit this structure could reduce the memory footprint substantially. At present we use full dense matrix diagonalization via the DSYEVR and DSYEVD subroutines from intel-lapack MKL. We retain this choice for now owing to the outstanding performance of these routines, even though it is not the optimal strategy for the sparse matrices generated by our pair list approach. Implementing on the fly diagonalizers adapted to this sparsity pattern is a natural next step.

\begin{acknowledgement}
The authors thank Rodrigo Neumann (IBM), Sumathy Raman (EMTech), Ramon Cardias (CBPF), Ivan S. Oliveira (CBPF) and Gustavo Petronilo (CBPF)  for support. This research was supported by ExxonMobil Technology and Engineering Company. Any opinions, findings, conclusions, or recommendations are those of the authors and not of the sponsors. 
\end{acknowledgement}
\section*{Code Availability}

The \vibra{} source code, documentation, and example input files are available
at Ref.~\cite{ViBraRepo}. The external proof-of-concept routines used for the
quantum vibrational spectroscopy workflow, including the modal-to-qubit
mapping, Pauli-operator construction, Hamiltonian fragmentation,
Trotterized time-evolution, and simulator-based sampling scripts, are available
at Refs.~\cite{QuantumVibMappingRepo,QuantumVibFragmentationRepo}. The calculations
reported in Section~\ref{sec:quantum_bridge} were performed using the code
versions identified by the repository release tags or commit hashes reported
in the corresponding references.
\section*{Data Availability}

The numerical data necessary to reproduce the validation tests, benchmark
calculations, tables, and figures reported in this work are provided in a
separate Zenodo data archive~\cite{QuantumVibData}. The archive includes the quartic force-field inputs, \vibra{} input files, reference spectra, selected VCI and symmetry-adapted VCI benchmark outputs, timing data, selected
configuration lists, and processed numerical data used to generate the figures
and tables. Data associated with the exploratory \ce{H2O} quantum-vibrational
proof-of-concept calculation, including sampled vibrational configuration
spaces and spectra generated from the external mapping, fragmentation, and
time-evolution routines, are also included. Source code and analysis scripts
are distributed separately, as described in the Code Availability statement.
\begin{suppinfo}

A comprehensive manual of \vibra{} is provided as Supporting
Information. It contains the detailed derivation of all equations, detailed
descriptions of every Fortran module and Python class, a complete input-keyword
reference, and an explanation of the output file formats. 
\end{suppinfo}

\bibliography{achemso-demo}

\end{document}